\documentclass[fleqn,usenatbib]{mnras}

\usepackage{newtxtext,newtxmath}

\usepackage[T1]{fontenc}

\DeclareRobustCommand{\VAN}[3]{#2}
\let\VANthebibliography\thebibliography
\def\thebibliography{\DeclareRobustCommand{\VAN}[3]{##3}\VANthebibliography}

\usepackage{graphicx}	
\usepackage{amsmath}	

\title[Future trajectories of the Solar System]
{Future trajectories of the Solar System: dynamical simulations of stellar encounters within 100 au}

\author[Raymond, Kaib, Selsis, Bouy]{
Sean N. Raymond,$^{1}$\thanks{E-mail: rayray.sean@gmail.com}
Nathan A. Kaib,$^{2}$
Franck Selsis,$^{1}$
and Herve Bouy$^{1}$
\\
$^{1}$Laboratoire d'Astrophysique de Bordeaux, CNRS and Universit{\'e} de Bordeaux, All{\'e}e Geoffroy St. Hilaire, 33165 Pessac, France \\
$^2$Planetary Science Institute, 1700 E. Fort Lowell, Suite 106, Tucson, AZ 85719, USA
}

\date{Accepted XXX. Received YYY; in original form ZZZ}

\pubyear{2023}

\begin{document}
\label{firstpage}
\pagerange{\pageref{firstpage}--\pageref{lastpage}}
\maketitle

\begin{abstract}
Given the inexorable increase in the Sun's luminosity, Earth will exit the habitable zone in $\sim 1$~Gyr. There is a negligible chance that Earth's orbit will change during that time through internal Solar System dynamics. However, there is a $\sim$~1\% chance per Gyr that a star will pass within 100~au of the Sun. Here, we use N-body simulations to evaluate the possible evolutionary pathways of the planets under the perturbation from a close stellar passage. We find a $\sim$~92\% chance that all eight planets will survive on orbits similar to their current ones if a star passes within 100~au of the Sun. Yet a passing star may disrupt the Solar System, by directly perturbing the planets' orbits or by triggering a dynamical instability. Mercury is the most fragile, with a destruction rate (usually via collision with the Sun) higher than that of the four giant planets combined. The most probable destructive pathways for Earth are to undergo a giant impact (with the Moon or Venus) or to collide with the Sun. Each planet may find itself on a very different orbit than its present-day one, in some cases with high eccentricities or inclinations. There is a small chance that Earth could end up on a more distant (colder) orbit, through re-shuffling of the system's orbital architecture, ejection into interstellar space (or into the Oort cloud), or capture by the passing star. We quantify plausible outcomes for the post-flyby Solar System.
\end{abstract}




\section{Introduction}

Earth has about a billion years of habitable surface conditions remaining.  Our planet is perched perilously close to the inner edge of the habitable zone~\citep{kasting93}.  If our planet were just 5-10\% closer to the Sun, there would no longer be a stable balance allowing for liquid water to remain on Earth's surface~\citep{selsis07,kopparapu13,leconte13}.  At the same time, the Sun's luminosity is slowly increasing, pushing the inner edge of the habitable zone inexorably outward, toward Earth's orbit~\citep[e.g.][]{wolf14}.  


Yet the Solar System is not a closed system. While the orbital evolution of the planets is largely determined by secular and resonant perturbations~\citep{laskar12}, passing stars can have a consequential influence on the planets' orbits. Based on the local galactic density of stars and their velocity dispersion, most stellar flybys are distant and only very weakly perturb the planets' orbits~\citep[e.g.][]{brown22}.  Statistically speaking, flybys closer than 100~au, which would strongly affect the planets' orbits, only take place roughly once per 100 Gyr in the current Galactic neighborhood~\citep{zink20,brown22}.  This amounts to a $\sim$~1\% probability of a sub-100~au encounter in the next Gyr, which is comparable to the well-studied probability of a chaos-driven dynamical instability among the terrestrial planets before the Sun becomes a red giant in $\sim 5$~Gyr~\citep{laskar09,zeebe15,abbot21}. 


Here, we use a suite of N-body simulations to evaluate the influence of a close stellar encounter on the orbital evolution of the Solar System. We explore the diversity of outcomes and calculate their relative probabilities. We pay particular attention to the future habitability of Earth, and show that there exist pathways by which Earth will end up on a cooler orbit (or a hotter one).  The most exotic outcomes involve capture of Earth or other planets by the passing star, or the scattering of planets into the Sun's Oort cloud. 

Our study was inspired in part by the pioneering work of \cite{laughlin00}, who simulated the dynamical response of the Solar System to flybys of binary stars.  We build on their results in a number of ways, for example, by including the Moon as a separate particle in a subset of simulations (Section 4), which allows for additional outcomes (such as the Moon crashing onto the Earth).  




\begin{figure*}
	\includegraphics[width=\columnwidth]{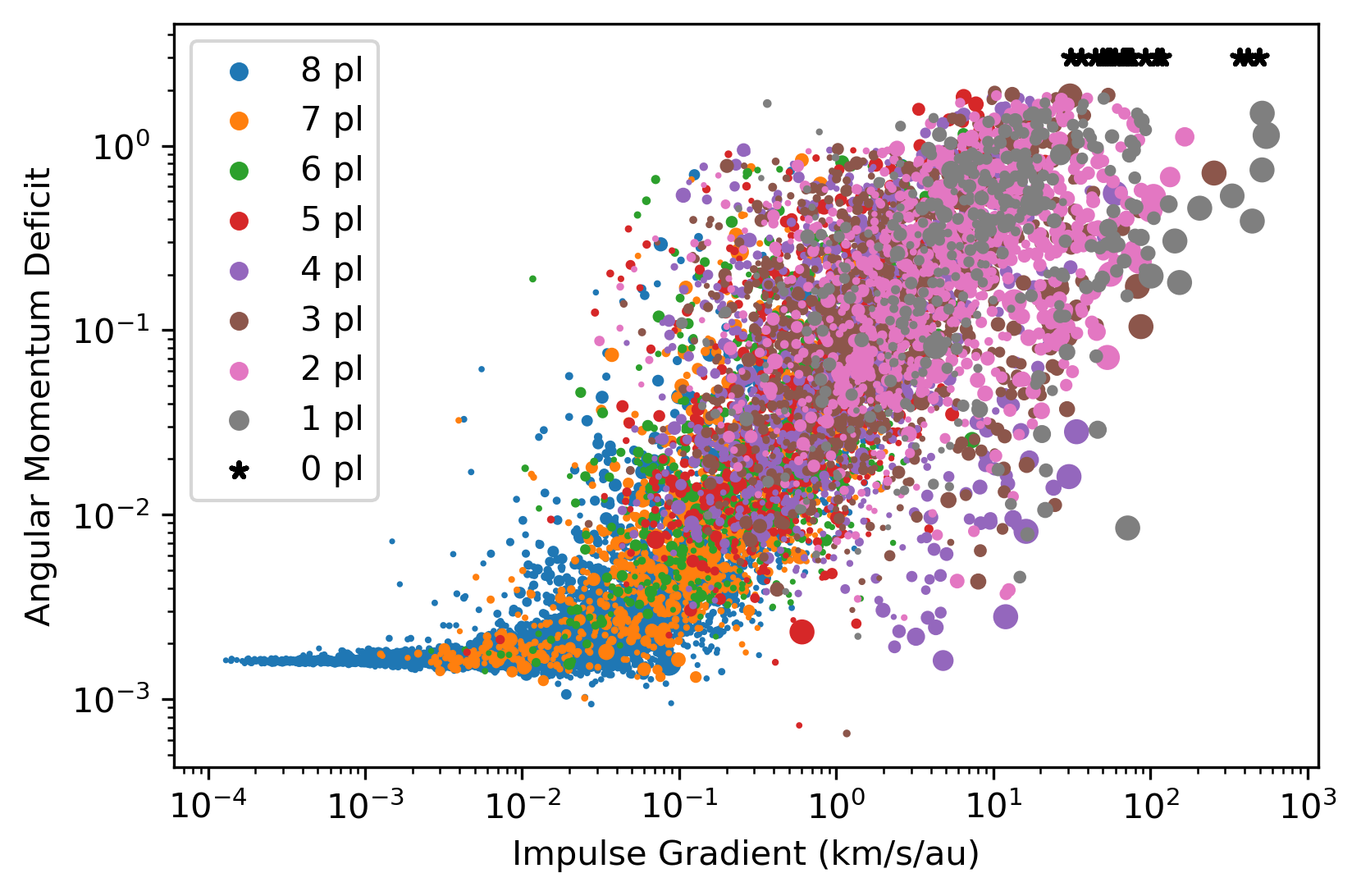}
	\includegraphics[width=\columnwidth]{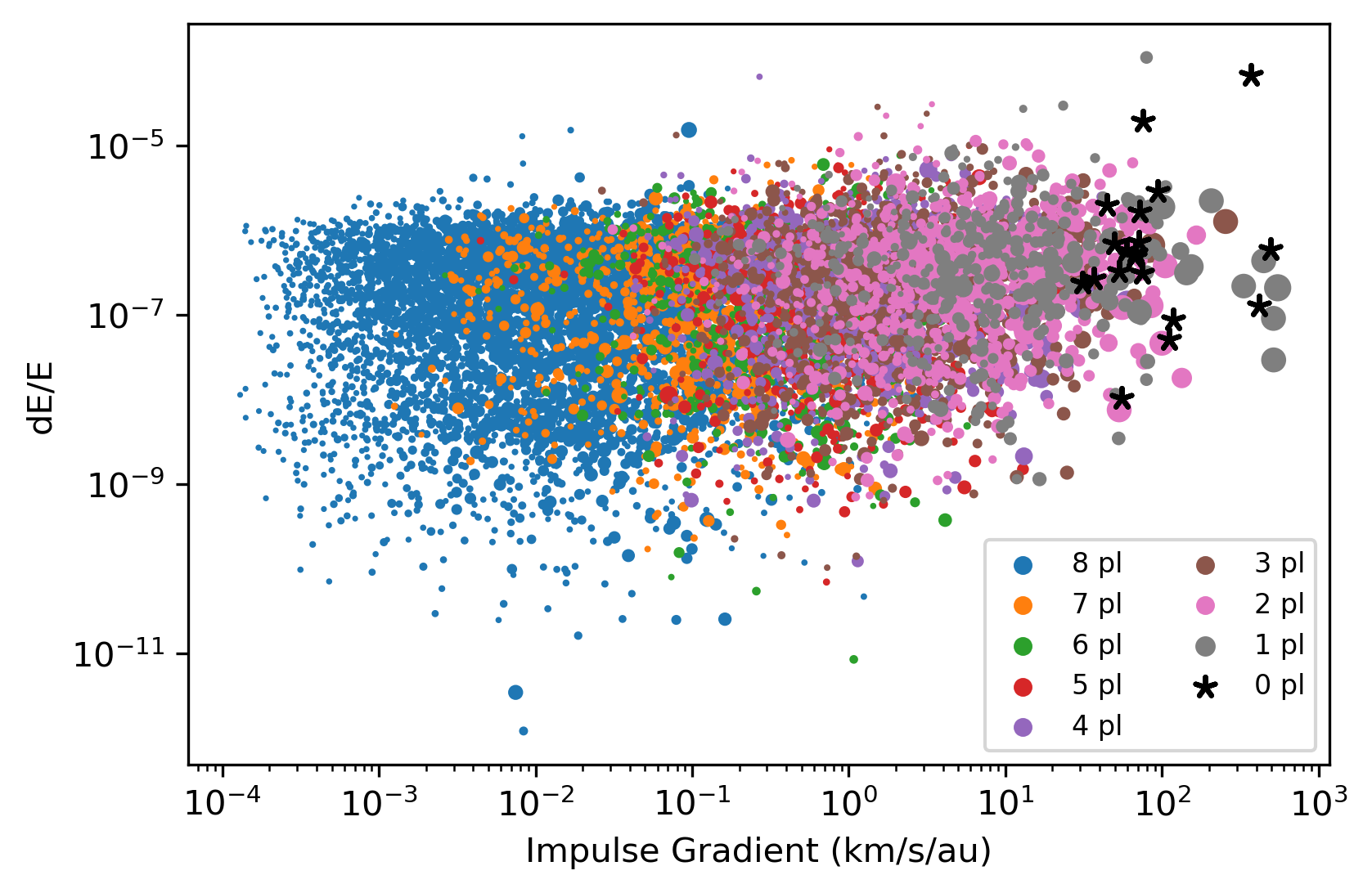}   
 \caption{Distribution of the outcomes of our simulations as a function of the impulse gradient of the stellar flyby.  The symbol color corresponds to the number of surviving planets in a given system, and the size is proportional to the mass of the flyby star. The angular momentum deficit $AMD$~\citep{laskar97,laskar17} is a strong function of the impulse gradient.  However, the relative energy conservation of individual simulations (as measured by the fractional energy error budget dE/E) is independent of the impulse gradient, lending credence to our numerical methods. }
    \label{fig:impgrad}
\end{figure*}

\section{Simulations}

\subsection{Initial conditions}
Our N-body simulations started with the eight planets orbiting the Sun, and a single star set to fly past.  The planets were placed at their positions as of Julian Date 2451000.5.\footnote{These are the planets' positions given as an example in the {\tt Mercury} code input files~\citep{chambers99}.}  The exact position at time zero has no influence on our results, because the timing and geometry of the closest stellar approach to the Sun varies from simulation to simulation between about 10 and 30 kyr due to differences in the stellar velocity.  The Earth and Moon were included as a single particle orbiting at their center of mass with the sum of their masses.  

Stellar encounters were generated to match the local Galactic neighborhood conditions. The flyby star's mass was chosen to follow the observed mass function of nearby stars determined by \cite{reid02}. The lower stellar mass limit was set to $0.05 M_\odot$, and any stars more massive than $4 M_\odot$ were set to $4 M_\odot$ to account for their short lifetimes. The orbital velocity of each star was chosen in a mass-dependent fashion to match the present-day velocity distribution in the Solar neighborhood~\citep[following][]{rickman08}. The resulting velocity distribution has a median of 40.2 km/s and a standard deviation of 16.9 km/s (averaged over all stellar types). We only include single stars.  At encounter speeds faster than $\sim 20$~km/s, which are typical given the velocity dispersion in the Galactic field, the Solar System's interaction with a binary can be approximated as separate interactions with each stellar component~\citep{li15b}.  This contrasts with the approach of \cite{laughlin00}, who only modeled binary encounters.

The final parameter for our stellar flybys was the impact parameter $b$. The impact parameter is essentially the closest approach of the flyby star before gravity is taken into account (bringing the closest approach closer). The probability of a close encounter scales as $b^2$, but we chose instead to sample following a $log(b)$ distribution. This choice allowed us to explore low-probability but interesting outcomes that arise from very close flybys.  We took this sampling into account when calculating probabilities (see Section 2.3).

The flyby star's orbit was placed at the edge of our numerical domain, just interior to the 100,000~au ejection radius.   While this is slightly smaller than the Sun's tidal radius~\citep[which depends on the local Galactic density;][]{tremaine93}, it still allows for capture of planets in the Oort cloud.

\subsection{Integration code}

Our code is based on the {\tt Mercury} integration package~\citep{chambers99}, to which we added two effects: general relativistic precession~\citep[using the implementation of][]{saha92}\footnote{See \cite{brown23} for an evaluation of the effect on Solar System stability of different implementations of general relativity.} and the Galactic tide~\citep[following][]{levison01b}.  In our calculation of the Galactic tide, we assumed a local galactic density of $0.1~\mathrm{M_\odot} \ pc^{-3}$~\citep{chakrabarti21}, and assigned values for Oort's constants of $\mathrm{A} = 14$~km/s/kpc and $\mathrm{B} = 12$~km/s/kpc~\citep{feast97}.\footnote{See \cite{raymond23c} for a study of the effect of varying the Galactic density on the trapping of scattered planets in the Oort cloud.}

In each of our 12000 simulations, the orbits of the planets and flyby star were integrated with a Bulirsch-Stoer integrator~\citep{press92}, using an accuracy parameter of $10^{-15}$.  Collisions were treated as inelastic mergers conserving linear momentum. Each simulation was integrated for 20 million years.

As a control, we also integrated the Solar System with no stellar flyby for 2 Gyr with the same code and saw no hints of instability.


\subsection{Probability determination}

As discussed above, our 12000 simulations sampled the impact parameters $b$ of stellar flybys using a logarithmic distribution instead of a physically-motivated, $b^2$ distribution.  We transform the probability of a simulation in our chosen sampling as follows.

Within our simulations, $d\mathrm{P_{sim}}/d (log (b))$ is constant, such that $d\mathrm{P_{sim}}/db \sim b^{-1}$.  In the real Universe, $d\mathrm{P_{Universe}}/db \sim b$.  To apply this to our simulations requires a linear distribution in $b$, so $d\mathrm{P_{Universe}}/db \sim b_{sim}/(d\mathrm{P_{sim}}/db)$, such that $d\mathrm{P_{Universe}}/db \sim b_{sim}^2$.

To evaluate the relative probability of our simulations, we therefore weight each one by its impact parameter squared.  We verified that this converges to the expected probability distribution for $b^2$ sampling. We normalize such that the sum of all of the probabilities is equal to one, keeping in mind that the total probability of a single encounter within 100~au is $\sim 1\%$ per Gyr given current local Galactic conditions~\citep{zink20,brown22}.

\section{Simulation outcomes}

\subsection{Influence of flyby parameters}

The key parameter in determining the outcome of a stellar flyby is the `impulse gradient', which measures the radial gradient in acceleration felt by planets orbiting the Sun. An impact parameter of 100~au and an encounter speed of 40 km/s yields an encounter timescale of ~12 years, or the orbital period of Jupiter. This implies that for such encounters, perturbations to the outer three giant planets can be approximated well within the impulsive regime. Although this is not necessarily true for Jupiter and the inner planets, we find that parameterizing our encounter results in the impulsive regime yields well-behaved empirical scalings. In particular, we find that encounter outcomes correlate well with the gradient of the stellar impulse measured at the Sun in the direction of the perpendicular to the stellar flyby path. The magnitude of this gradient is defined as:
\begin{equation}
    \text{Impulse Gradient} = \frac{2 G M_\star}{v_\star b^2}, 
\end{equation}
\noindent where $M_\star$ is the mass of the flyby star, $v_\star$ is its speed relative to the Sun (at infinity), $b$ is the impact parameter of the flyby, and $G$ is the gravitational constant.  The units of the impulse gradient are generally s$^{-1}$, but we express them in km/s/au, to give a more intuitive sense of the perturbations felt by the planets, as their orbital speeds are measured in tens of km/s.  

The normalized angular momentum deficit is a measure of the orbital excitation level of a planetary system, in terms of the non-circularity and non-coplanarity of the planets' orbits~\citep{laskar97}.  It is defined as:
\begin{equation}
 AMD = \frac{\sum_{j} m_j \sqrt{a_j} \left(1 - cos
(i_j) \sqrt{1-e_j^2}\right)} {\sum_j m_j \sqrt{a_j}}, 
\end{equation}
\noindent where $a_j$, $e_j$, $i_j$, and $m_j$ refer to planet $j$'s semimajor axis, eccentricity, inclination with respect to a fiducial plane, and mass. The $AMD$ of the Solar System is 0.00128 (or 0.0015 when measured relative to the Earth's orbital plane).

Figure~\ref{fig:impgrad} (left panel) shows that the angular momentum deficit of surviving systems is a strong function of the impulse gradient in our simulations. For weak stellar encounters (with impulse gradients below $\sim 10^{-2}~$km/s/au), the Solar System is barely perturbed by the passing star, and the planets' orbits almost always maintain their same low-eccentricity, low-inclination orbits. A planet is lost in only 3.4\% of simulations with $IG < 10^{-2}$~km/s/au (although, when weighted by the relative probability of each encounter, only 1.4\% of encounters with $IG < 10^{-2}$~km/s/au lose a planet). These encounters are not universally at large distances; while the median impact parameter is 47.5~au among flybys with $IG < 10^{-2}$~km/s/au, a handful have $b < 10$~au and one quarter of stars crossed the planets' orbits, with $b < 30$~au.  For stronger encounters, with $IG > 10^{-2}$~km/s/au, the strength of the instability still scales with the impulse gradient.  The level of devastation can be measured in terms of both the $AMD$ and simply by the number of surviving planets.  In the most extreme cases, one (or no) planets survive.  We will explore the evolution of such cases below.

The degree of orbital energy conservation $dE/E$ is virtually independent of the impulse gradient (Fig.~\ref{fig:impgrad}, right panel).  This is reassuring, as it would be concerning if the strength of the instability governed the level of accuracy of our simulations.  In addition, the overall level of energy conservation is excellent, with a median of $1.9 \times 10^{-7}$, which is orders of magnitude lower than the acceptable threshold used in simulations of giant planet scattering~\citep[e.g.][]{chatterjee08,raymond10}.  This demonstrates that our integration method is reliable. 

\subsection{Number of surviving planets}
\begin{figure}
	\includegraphics[width=\columnwidth]{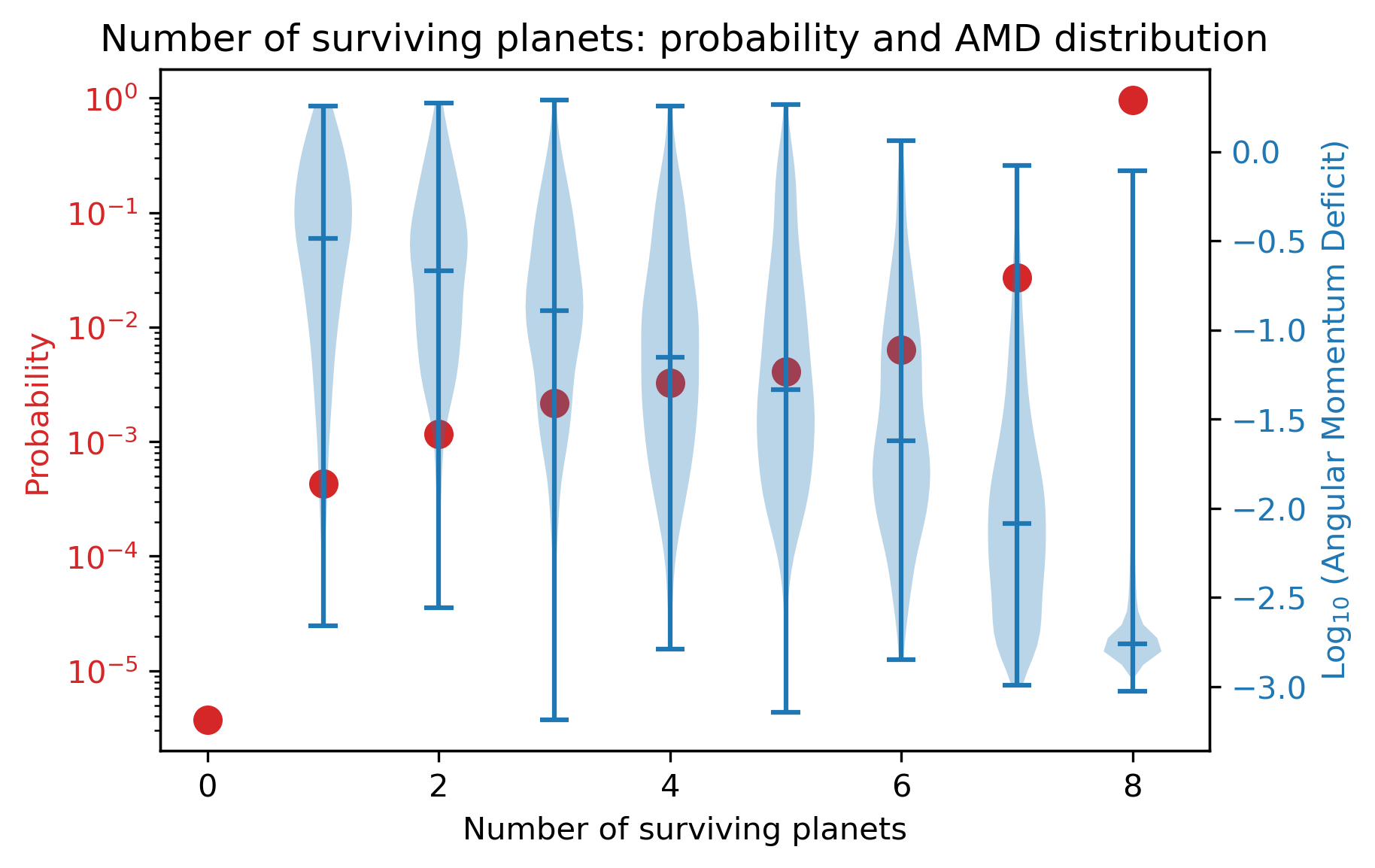}
    \caption{Probability of $N_p$ planets surviving a stellar flyby within 100~au (left axis; red). The right (blue) axis shows the angular momentum deficit distribution as a function of $N_p$. The central tick shows the median of each distribution, and the other ticks show the maxima and minima of the $AMD$ distribution for each value of $N_p$.  The present-day Solar System's $AMD$ is $\sim$0.00157, or -2.8 in this plot.}
    \label{fig:np_hist}
\end{figure}

If a star passes within 100~au of the Sun, there is still a very high chance that all 8 Solar System planets will survive.  Figure~\ref{fig:np_hist} shows a probability of 95.6\% that no planets will be lost, at least during the 20 Myr following the stellar flyby. There is a 92\% probability that all 8 planets will survive with each pair of neighbors on Hill-stable orbits with separations larger than 3.5 mutual Hill radii~\citep{marchal82,gladman93}, and the system will have an angular momentum deficit less than twice the present-day value.  There is a monotonically decreasing probability of fewer planets surviving, with only a $3.7\times 10^{-6}$ probability of zero remaining planets.

The angular momentum deficit $AMD$ of post-flyby systems scales strongly with the number of surviving planets $N_p$ (Fig.~\ref{fig:np_hist}).  For systems with 8 surviving planets, the median $AMD$ is only 10\% higher than the present-day Solar System's. The orbits of systems with fewer surviving planets are progressively more and more excited as $N_p$ decreases, with higher and higher eccentricities.  This same trend is seen among systems of known exoplanets~\citep{limbach15,turrini20}.  

\subsection{Examples}

\begin{figure*}
	\includegraphics[width=0.66\columnwidth]{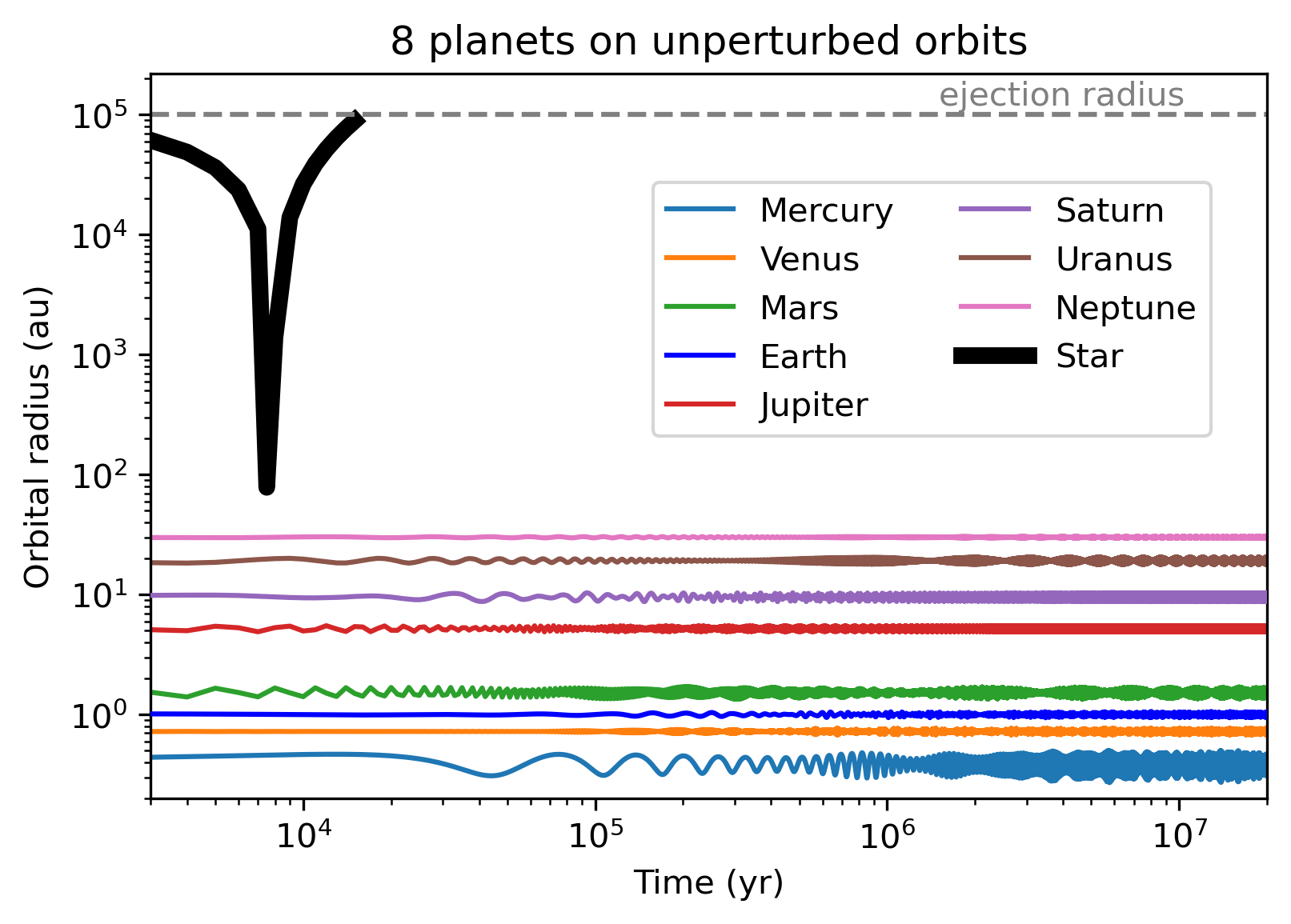}
 	\includegraphics[width=0.66\columnwidth]{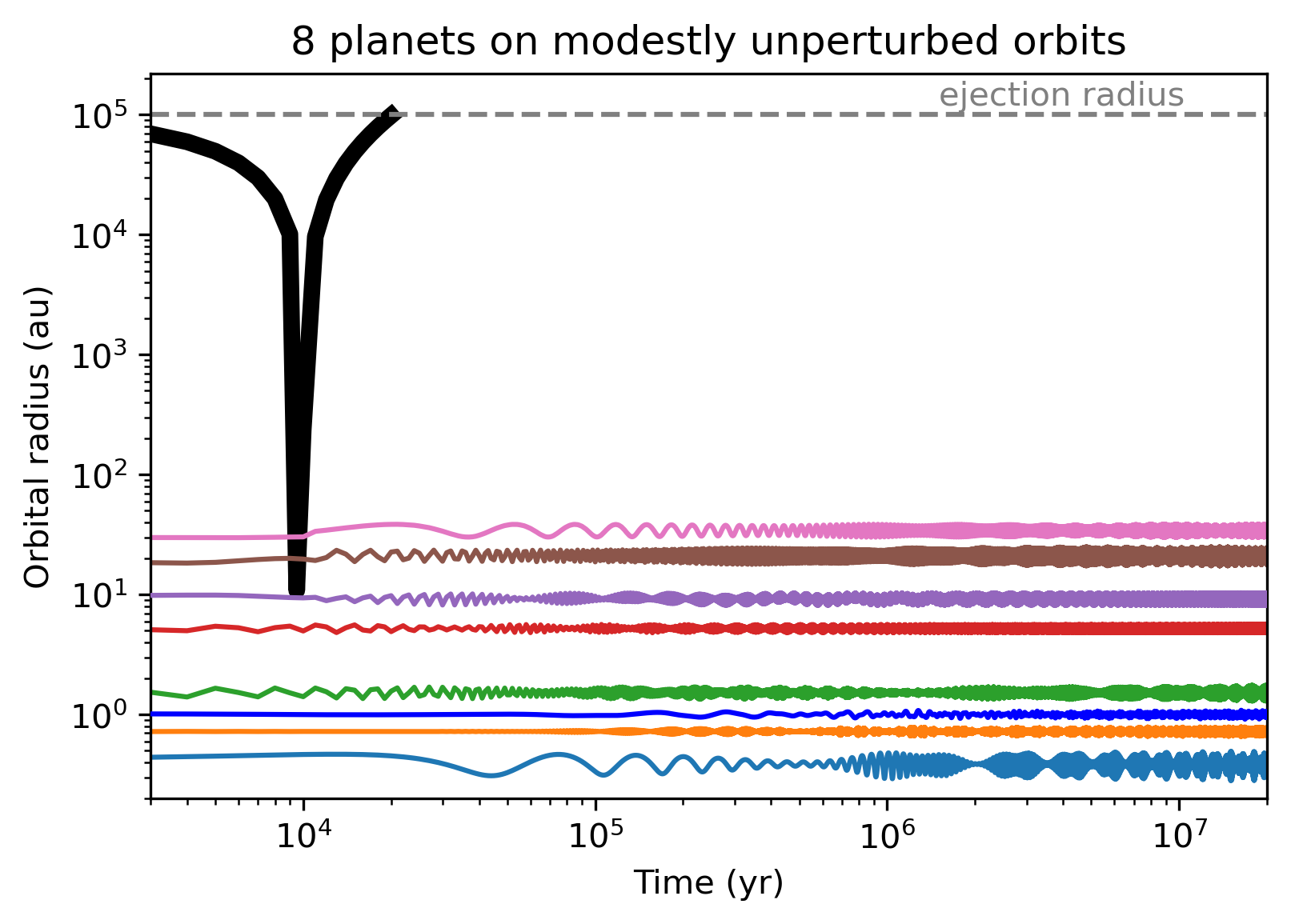}
	\includegraphics[width=0.66\columnwidth]{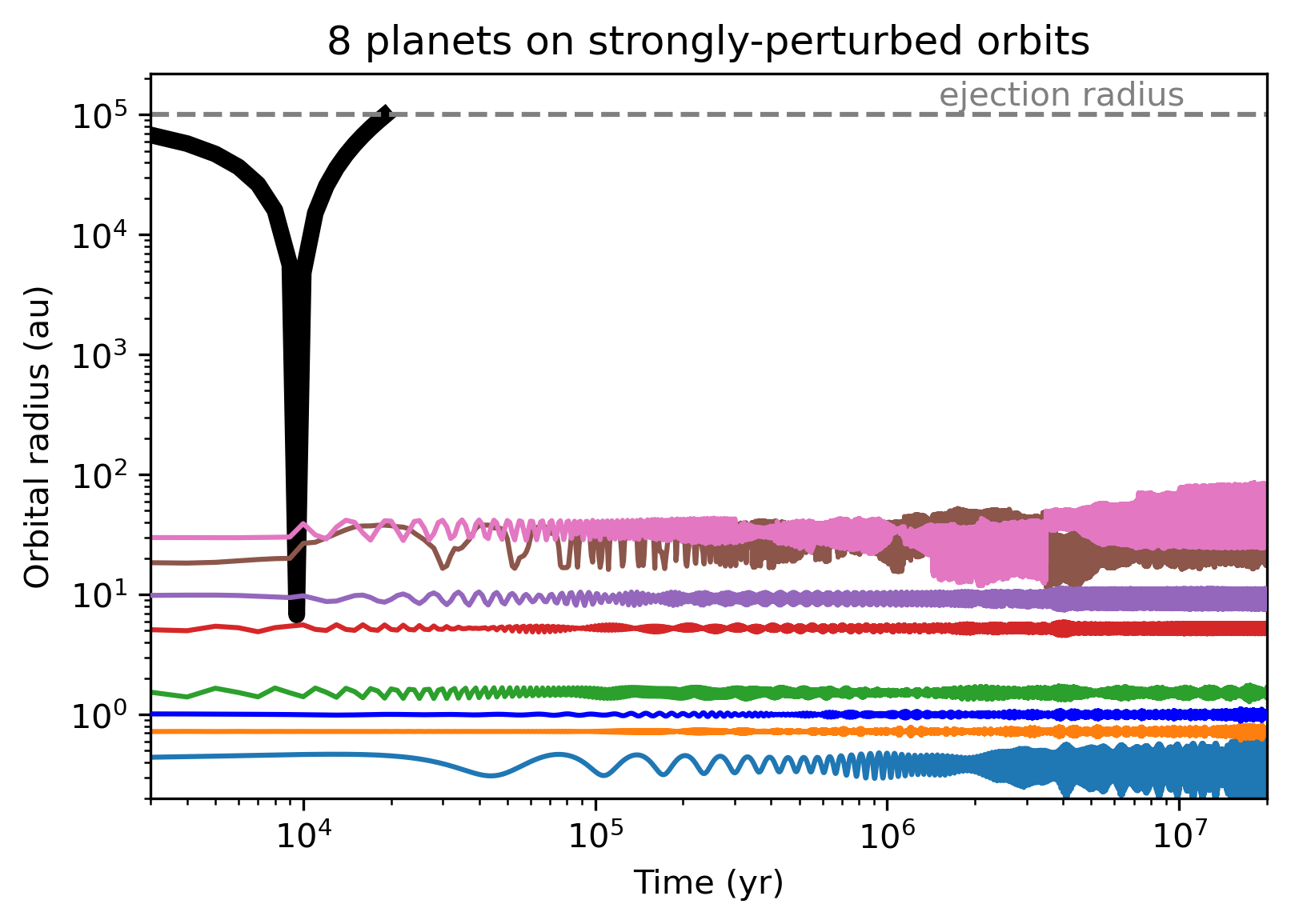}\\
	\includegraphics[width=0.66\columnwidth]{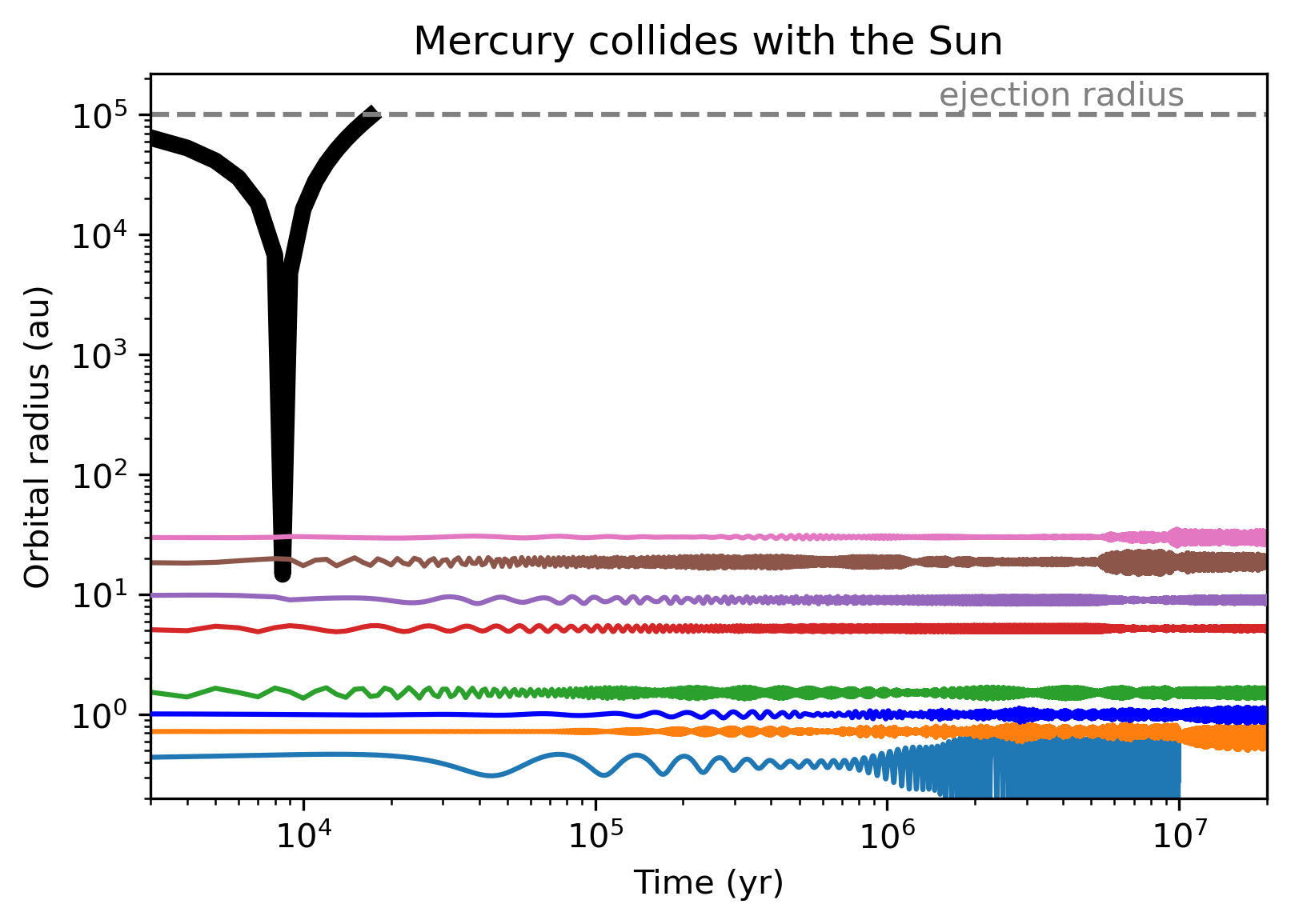}
	\includegraphics[width=0.66\columnwidth]{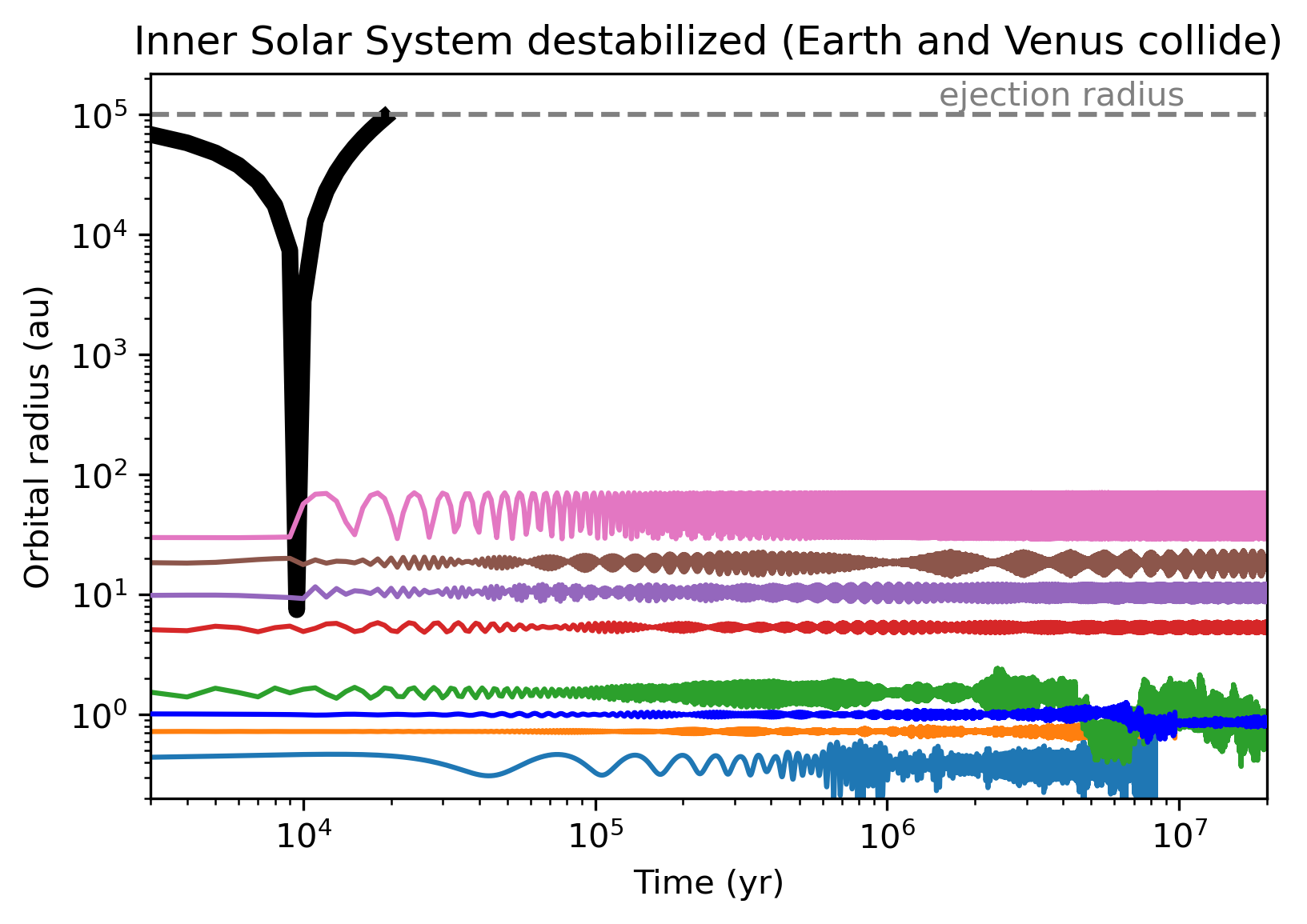}
	\includegraphics[width=0.66\columnwidth]{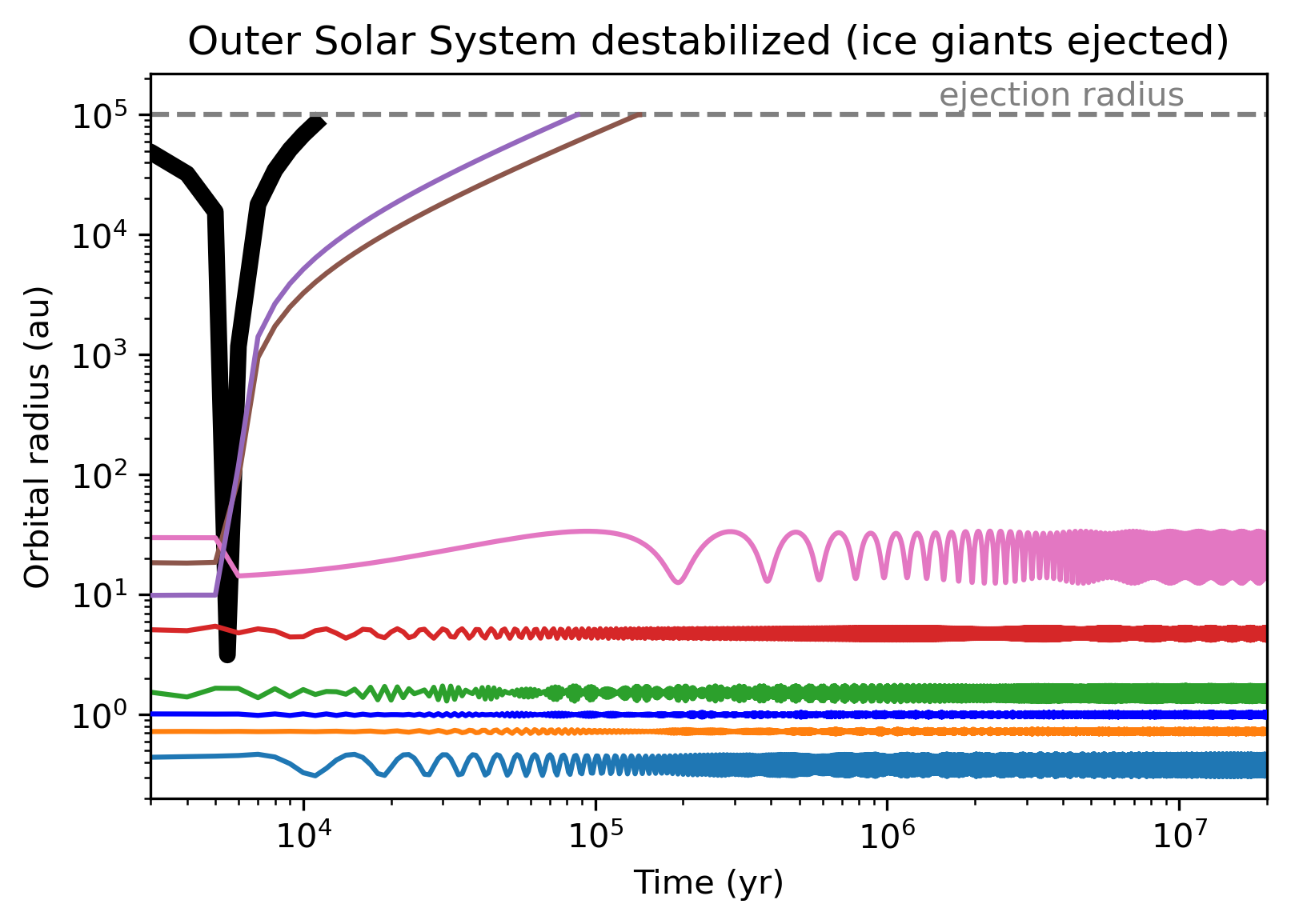}\\
	\includegraphics[width=0.66\columnwidth]{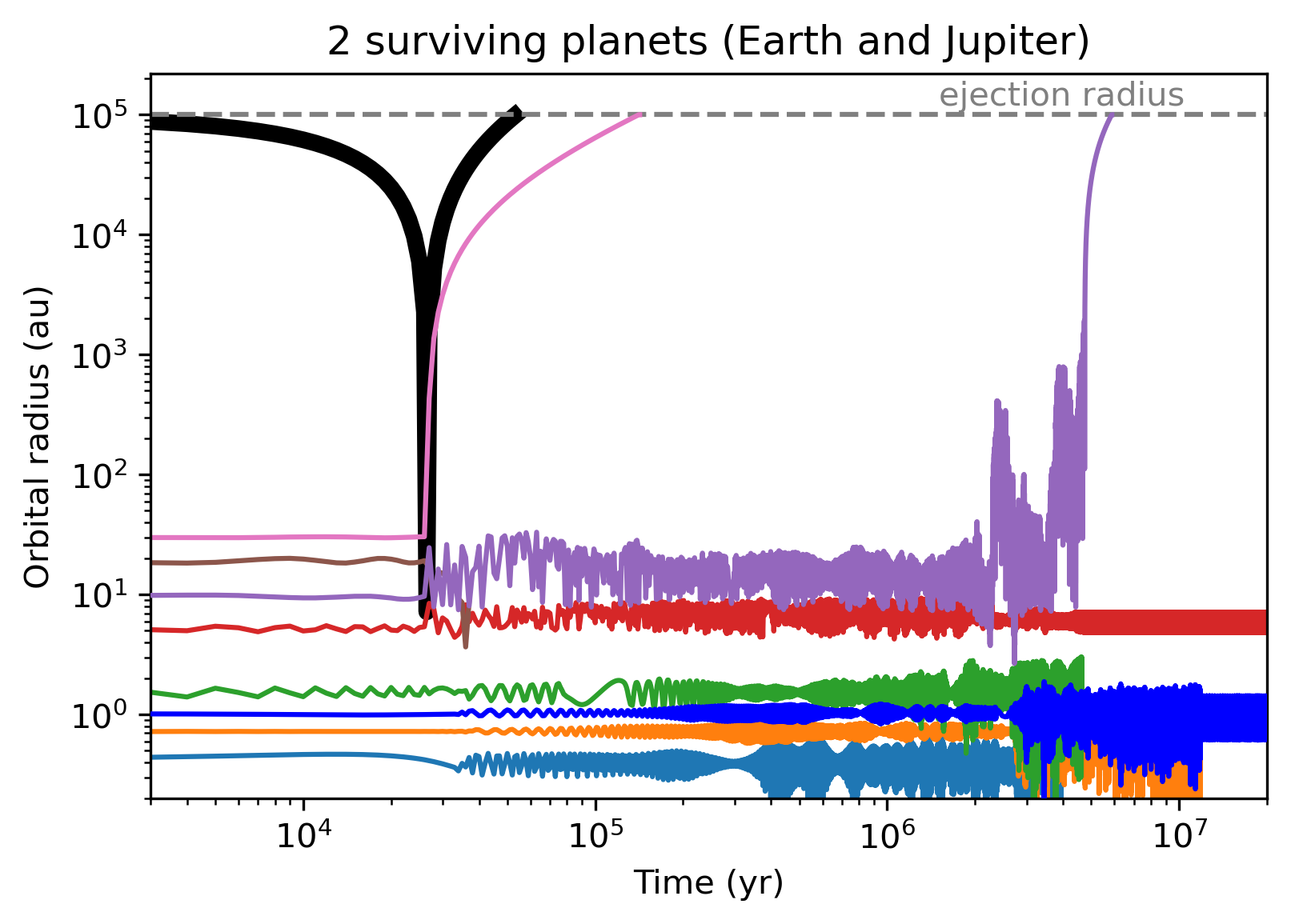}
	\includegraphics[width=0.66\columnwidth]{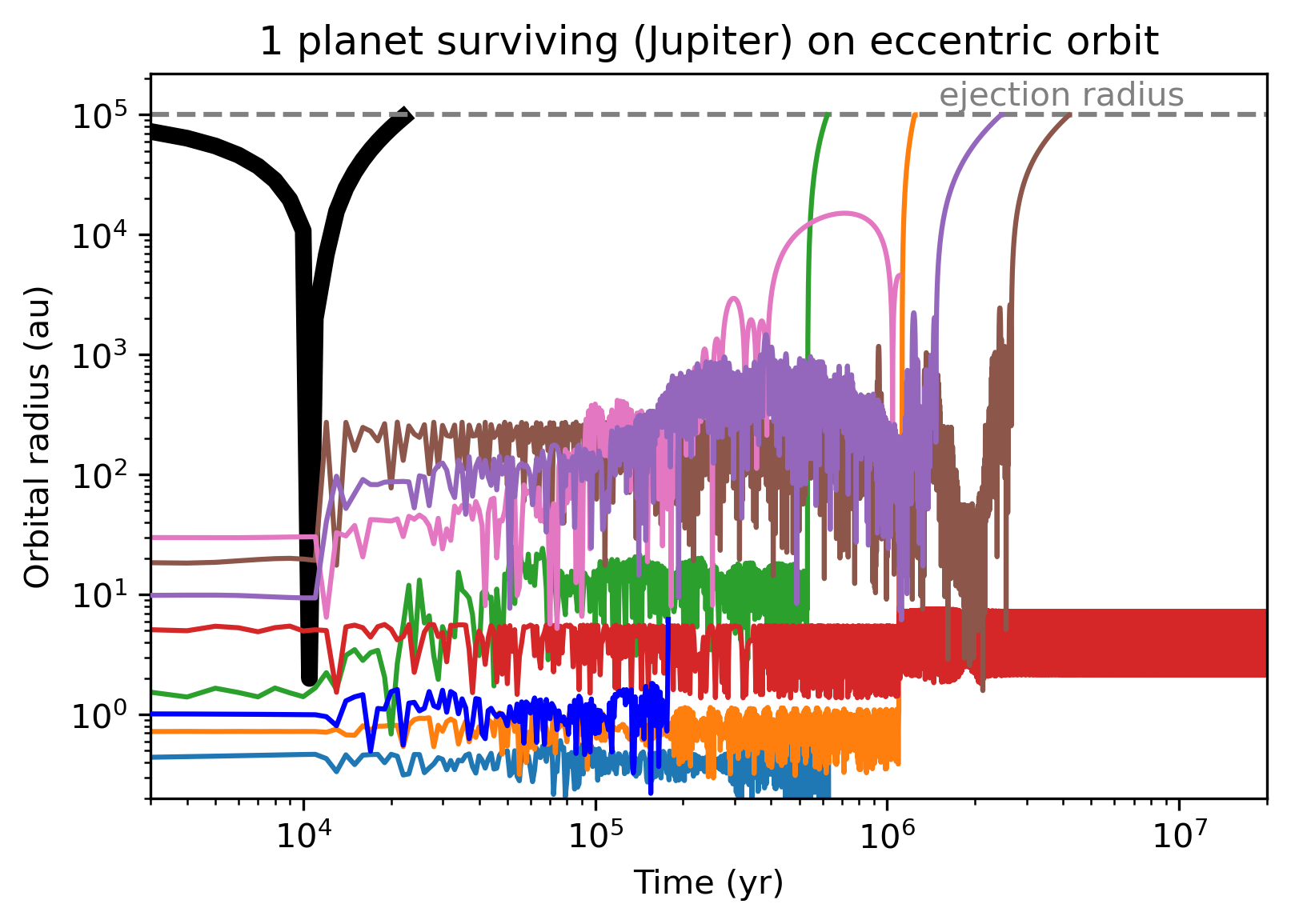}
	\includegraphics[width=0.66\columnwidth]{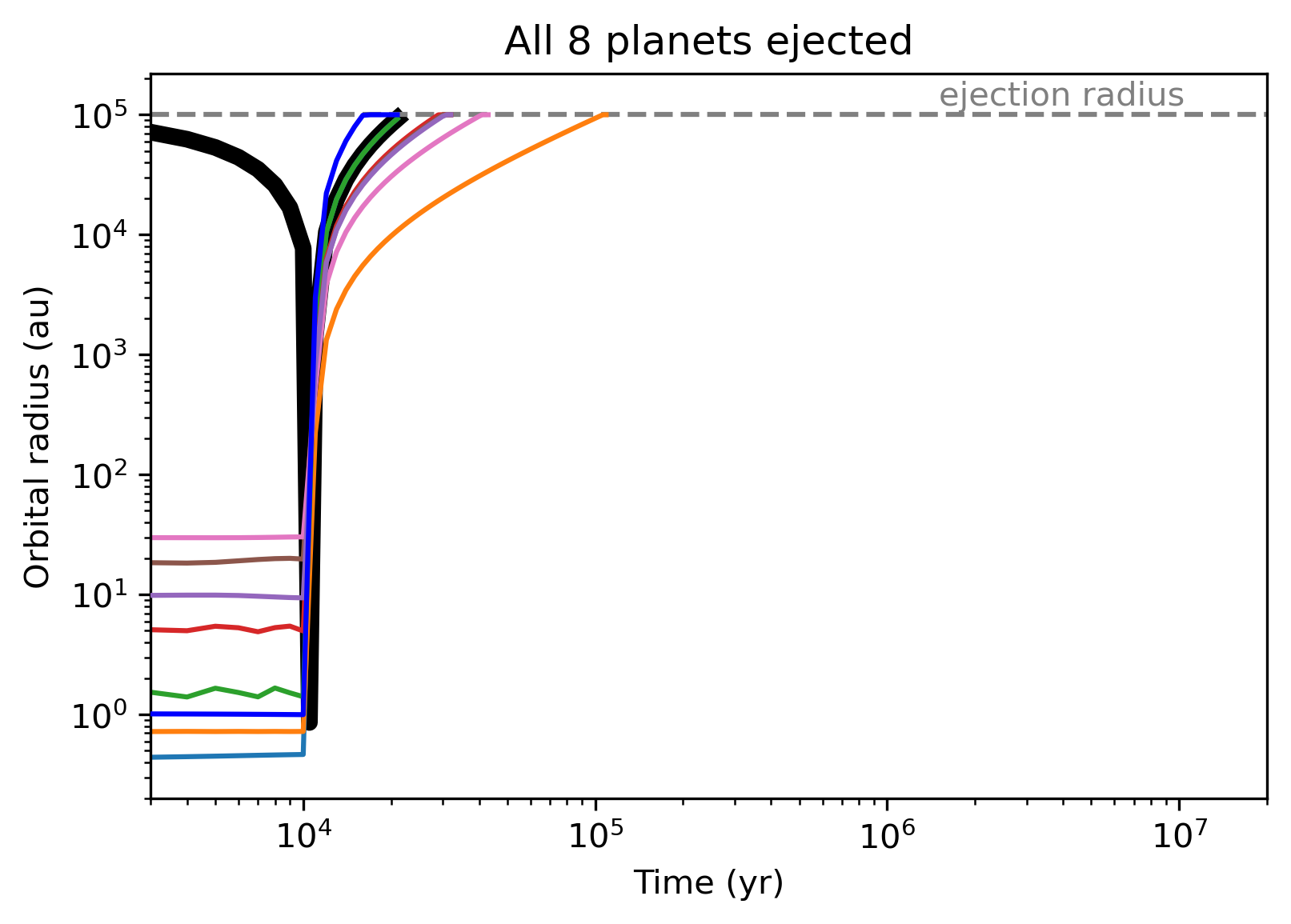}
    \caption{Nine example simulations illustrating the diversity of outcomes.  Each panel shows the orbital radius of each planet and the flyby star as a function of time; note that this is not the semimajor axis, so the spread in a planet's orbital radius over time is a measure of its orbital eccentricity. The top row includes three simulations in which all eight planets survived; the top left simulation is basically unperturbed, while the top right simulation is in the early stages of dynamical instability.  The middle row includes systems that lost 1-2 planets.  The bottom row presents simulations that were catastrophically disrupted, with 0-2 surviving planets. }
    \label{fig:examples}
\end{figure*}

Figure~\ref{fig:examples} shows the evolution of nine example simulations that span the range of outcomes.  The top row shows three cases in which all of the planets survived, but with different levels of orbital excitation.  The simulation at the top left underwent a weak encounter and had a final $AMD$ that was indistinguishable from its starting value.  The top center simulation underwent a stronger instability and ended up with an $AMD$ about three times higher than the pre-encounter Solar System.  However, the planets' orbits remained stable with no orbital crossing.  In the top right simulation, a still stronger instability perturbed the system to a very high $AMD$, with clear signs of impending instability, as Uranus and Neptune's orbits are crossing and Mercury's eccentricity has drastically increased.  

The center row of Fig.~\ref{fig:examples} shows cases in which one or more planets were lost from the Solar System.  In the center left simulation, perturbations to the giant planets were transmitted to the terrestrial planets, leading Mercury to eventually collide with the Sun after $\sim$~10 Myr.  In the center middle simulation, a stronger stellar encounter significantly altered the ice giants' orbits and ended up destabilizing the terrestrial planets.  Mercury once again collided with the Sun, and Venus and Earth underwent a giant impact.  In the center right panel, the stellar perturbations were largely confined to the outer Solar System: Uranus and Saturn were both ejected from the Solar System but the remaining planets' orbits were stable for the rest of the simulation.

\begin{figure}
	\includegraphics[width=\columnwidth]{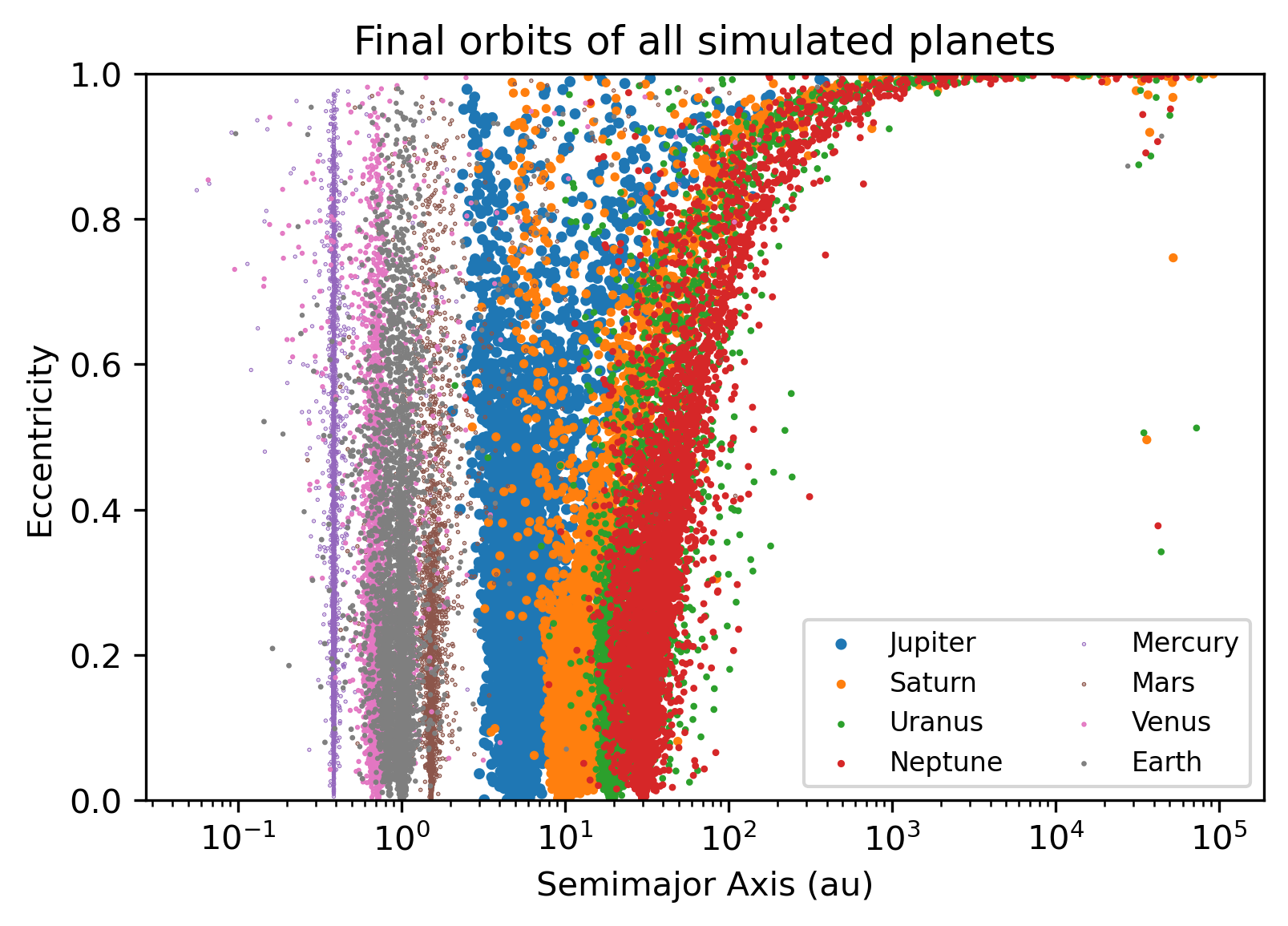}
    \caption{Final orbits of all surviving planets in our simulations. The tail of planets with semimajor axes of $10^{4-5}$~au are those trapped in the Oort cloud.}
    \label{fig:ae_all}
\end{figure}

The bottom row of Fig.~\ref{fig:examples} shows a few of the most destructive outcomes.  In the bottom left simulation, the stellar encounter rapidly drove Uranus into the Sun, ejected Neptune, and triggered a dynamical instability between Saturn and Jupiter.  Saturn was eventually ejected by Jupiter, and the gas giant perturbations spread to the terrestrial planets, leading to Mercury, Venus and Mars colliding with the Sun.  The final Solar System contained just Jupiter and Earth, on eccentric orbits ($e_{Earth} \approx 0.4, e_{Jup} \approx 0.2$).  The bottom center simulation shows a case in which only Jupiter survived.  The very strong stellar encounter (with $b = 2.6$~au) immediately triggered a strong dynamical instability across the entire Solar System.  Mercury, Earth and Neptune collided with the Sun.  Venus, Mars, Saturn, and Uranus were ejected after close encounters with Jupiter.  Jupiter was the lone survivor, winding up with an orbital eccentricity of 0.55.  Finally, in the bottom right simulation from Fig.~\ref{fig:examples}, an extremely strong close encounter with an impact parameter of 1.5~au and an impulse gradient of $2\times 10^{-7}$ caused all 8 planets to be ejected, leaving the Sun planet-less and alone.


\subsection{Post-flyby orbits}

Figure~\ref{fig:ae_all} shows the final (post-flyby) orbital distribution of the planets in all of our simulations.  Each planet has a distribution of final orbits that extends to indefinitely high orbital eccentricities (up to $e \sim 1$).  Their semimajor axis-eccentricity distribution is reminiscent of those of the survivors of giant planet instabilities such as those thought to be responsible for the large eccentricities observed among giant exoplanets~\citep{chatterjee08,raymond10}.  A key difference is that, in our simulations, only the giant planets have high enough escape speeds to eject other planets~\cite[or, strictly speaking, large enough Safronov numbers $\Theta$, where $\Theta$ is the ratio of the escape speed from a planet's surface to the escape speed from the system at the planet's orbital radius; see][]{ford08}. Instabilities among the terrestrial planets usually lead to collisions, not ejections.  Yet perturbations from the giant planets can clearly leave the terrestrial planets on orbits that are excited well beyond their self-excitation limit.

\begin{figure*}
	\includegraphics[width=\columnwidth]{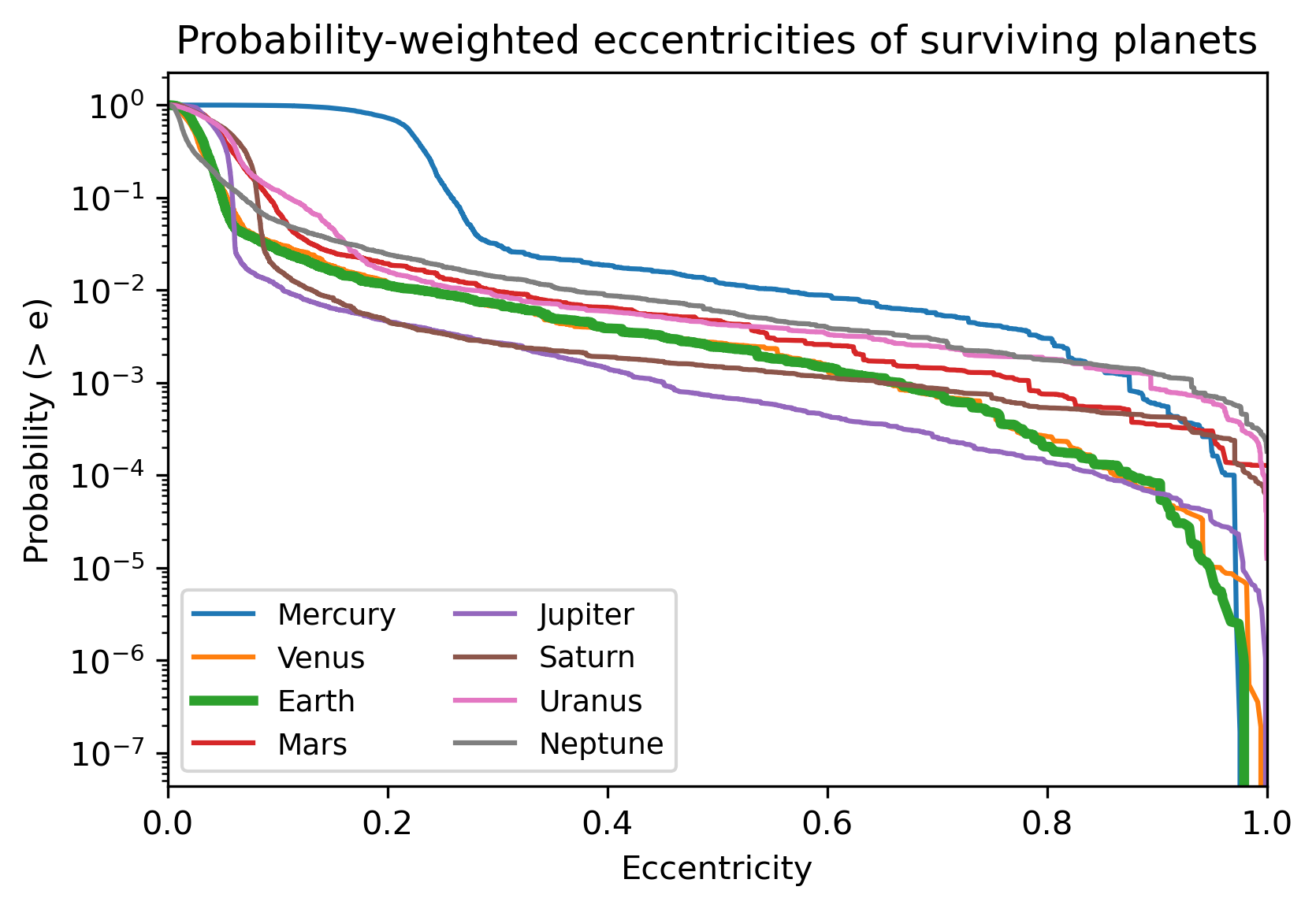}
	\includegraphics[width=\columnwidth]{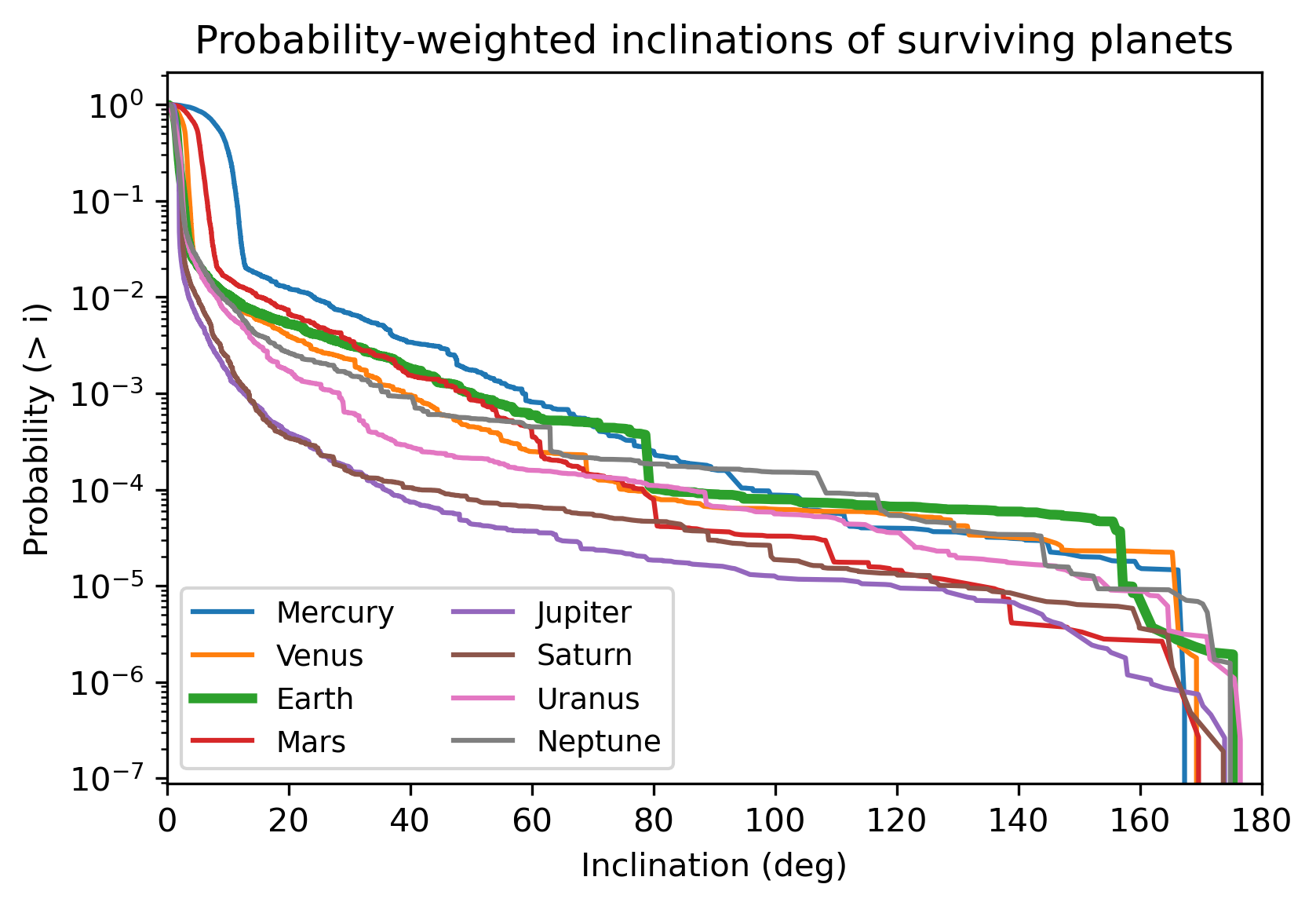} 
    \caption{Eccentricity (left) and inclination (right) distributions of surviving planets, weighted by the probability of each outcome.  Each plot shows a reverse cumulative distribution; the y-axis shows the probability that a given planet has an eccentricity or inclination larger than a specified value. }
    \label{fig:ei_dist}
\end{figure*}

Figure~\ref{fig:ei_dist} shows the probability-weighted eccentricity and inclination distributions for all 8 planets (when they survived).  The distribution is strongly weighted toward simulations in which the Solar System remained stable and with a low $AMD$, with the planets on orbits similar to their present-day ones, because that is the most probable outcome of a sub-100~au flyby (see Fig.~\ref{fig:np_hist}).  However, there is a tail for each planet that extends to high eccentricities and inclinations.  These correspond to dynamical instabilities triggered by the flyby that are violent enough to strongly excite the planets' orbits.  Of course, when instabilities are so strong that planets are ejected from the Sun, they no longer contribute to the distributions in Fig.~\ref{fig:ei_dist}.  While low in overall probability, outcomes in which the planets survive on high eccentricity or highly-inclined orbits are intriguing in terms of their influence on planetary climate.  We will return to this issue in Section 5, with regards to the Earth. 

A striking population of surviving planets in Fig.~\ref{fig:ae_all} are those with semimajor axes of $10^{4-5}$~au but small enough eccentricities that they are not in dynamical contact with the other planets.  These are Oort cloud planets~\citep[see][which is entirely dedicated to Oort cloud planets]{raymond23c}.  They were scattered outward during gravitational instabilities, and spent long enough at very high orbital radius to have had their perihelion distances increased via torques from the Galactic tide, until they were no longer being scattered by the inner planets.  As such, their origins are similar to those of Oort cloud comets~\citep[e.g.][]{brasser13b,dones15}.  

\begin{figure}
	\includegraphics[width=\columnwidth]{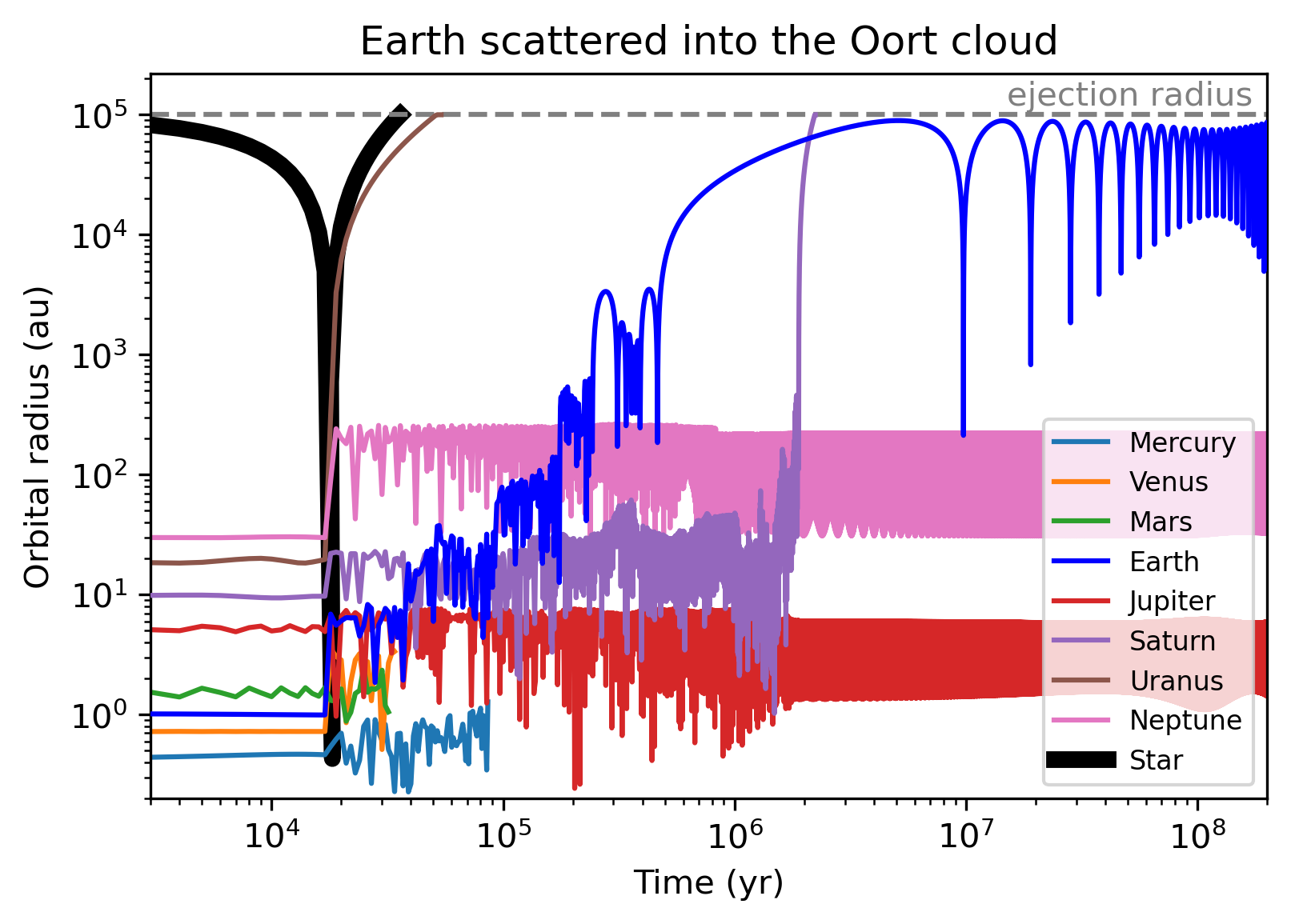}
    \caption{A simulation in which Earth was scattered out by the gas giants and captured in the Oort cloud. }
    \label{fig:oort_e}
\end{figure}

We performed additional simulations to capture the implantation of Oort cloud planets.  Starting from the final orbits after 20 Myr, we selected candidate systems in which a planet's semimajor axis was larger than 1000~au and its perihelion distance was larger than 50~au.  We integrated these systems out to 200 Myr. In many cases the planet was simply ejected, but in 28 simulations a planet was captured in the Oort cloud.  In most cases, it was Saturn or the ice giants that were captured in the Oort cloud, but there was one in which Mars ended up in the Oort cloud, and there were two simulations in which Earth was captured as an Oort cloud planet.  The total probability of a planet being captured in the Oort cloud in our simulations was $8.9 \times 10^{-5}$, which, after taking into account the encounter likelihood, amounts to a net probability of $8.9 \times 10^{-7}$ per Gyr.  This is likely an underestimate; at the end of our 20 Myr simulations there were an additional 182 planets with semimajor axes larger than 1000~au and perihelion distances smaller than 50~au.  Assuming an Oort cloud trapping rate of $\sim$5\% based on previous results~\citep{raymond23c}, this would increase the total probability of trapping a planet in the Oort cloud by roughly 30\%, to $1.2 \times 10^{-6}$ per Gyr.

Figure~\ref{fig:oort_e} shows the evolution of a simulation in which Earth was trapped in the Oort cloud.  After being scattered by the giant planets to large orbital radius the Galactic tide increased the Earth's perihelion distance on a $\sim 100$~Myr timescale.  Earth finished the simulation on a stable orbit in the Oort cloud with an orbital semimajor axis of 54977~au.  The long-term survival of Earth in the Oort cloud is not guaranteed. Torques from the Galactic tide continue to affect the Oort cloud planet's orbit, eventually decreasing its perihelion distance and bringing it back into dynamical contact with the inner planets (Neptune and Jupiter in this case). Earth could potentially be ejected from this simulation on a timescale of a few hundred million years~\citep[see][]{raymond23c}.  Also, additional stellar encounters are likely to strip a 50,000~au-wide orbit after a Gyr or so~\citep[see Fig 18 of][]{kaib11b}.

\subsection{Planetary destruction pathways}

\begin{table*}
	\centering
	\caption{Probabilities of outcomes for each planet if a star passes within 100~au of the Sun -- these should be multiplied by 0.01 per Gyr for absolute probabilities}
	\label{tab:outcomes}
	\begin{tabular}{l|c|c|c|c|c} 
		\hline
		Planet & Collision with Sun & Impact with other planet & Ejection & Capture by flyby star & Trapped in Oort cloud\\
		\hline
		Mercury & 0.025 & 0.0080 & $6.2 \times 10^{-4}$ & $4.0 \times 10^{-6}$ & -- \\
		Venus & 0.0033 & 0.012 & $6.0 \times 10^{-4}$ & $1.2\times 10^{-5}$ & --\\
		Earth & 0.0024 & 0.0048 & $5.8 \times 10^{-4}$ & $1.1\times 10^{-5}$ & $2.1\times 10^{-6}$ \\
		Mars & 0.012 & 0.0027 & $8.6 \times 10^{-4}$ & $2.0\times 10^{-5}$ & $1.7\times 10^{-6}$ \\
		Jupiter & $9.2\times 10^{-6}$ & $5.5 \times 10^{-4}$ & $5.5 \times 10^{-4}$ & $2.4\times 10^{-5}$ & -- \\
		Saturn & $5.0 \times 10^{-4}$ & $3.5 \times 10^{-4}$ & 0.0032 & $4.0\times 10^{-5}$ & $1.0\times 10^{-5}$ \\
		Uranus & 0.0017 & $5.4 \times 10^{-4}$ & 0.011 & $4.6\times 10^{-5}$ & $1.5\times 10^{-5}$ \\
		Neptune & $7.2 \times 10^{-4}$ & $2.9 \times 10^{-4}$ & 0.0081 & $5.4\times 10^{-6}$ & $6.3\times 10^{-5}$\\

		\hline
	\end{tabular}
\end{table*}

We determined the most common pathways through which planets may be lost, keeping in mind that there is a $\gtrsim 95\%$ chance that no planet will be lost if a star passes within 100~au (see Fig.~\ref{fig:np_hist}).  Table~\ref{tab:outcomes} shows these probabilities for each planet.

The ten most likely destruction planetary pathways are, in order of decreasing probability:
\begin{enumerate}
    \item Mercury collides with the Sun (probability of 2.54\%).
    \item Mars collides with the Sun (1.21\%).
    \item Venus impacts another planet (1.17\%).
    \item Uranus is ejected (1.06\%).
    \item Neptune is ejected (0.81\%).
    \item Mercury impacts another planet (0.80\%).
    \item Earth impacts another planet (0.48\%).
    \item Saturn is ejected (0.32\%).
    \item Mars impacts another planet (0.27\%).
    \item Earth collides with the Sun (0.24\%).
\end{enumerate}
Examples of many of these outcomes are apparent in Fig.~\ref{fig:examples}. These outcomes are not mutually exclusive; in many cases multiple planets are lost in the same simulation.  

Mercury's destruction rate is higher than that of all the giant planets put together.  It is the most fragile planet in the Solar System from a dynamical point of view, even through perturbations from passing stars. It is already well-established that, in the absence of a stellar flyby, Mercury is the most likely planet to be lost from the Solar System, by having its eccentricity dramatically increased via a secular resonance with Jupiter, causing it to collide with the Sun~\citep{laskar09,batygin15b,zeebe15,abbot21}.  Mercury's eccentricity increase can also result in collisions between the rocky planets, which are also common outcomes in simulations with stellar flybys.  In such simulations, Mercury's destruction does not happen until long after the stellar flyby.  We interpret this as evidence that its eccentricity increase is induced by indirect perturbations, through secular forcing by the giant planets -- likely via the same secular resonance with Jupiter that is capable of destabilizing its orbit in the absence of flybys.

It is not surprising that the ice giants are the most susceptible to being ejected, as they are more loosely bound to the Sun than the terrestrial planets, inhabiting the realm of parameter space in which ejections are favored over collisions~\citep[e.g.][]{ford08}, and they are particularly vulnerable to gravitational scattering by Jupiter. Saturn, on the other hand, was typically only ejected after an ice giant's orbit was destabilized (and likely already ejected), bringing it into dynamical contact with Jupiter.

A single planet was destroyed in 1130 of our 12000 simulations. The total probability of losing just one planet if a star passes within 100~au is 2.7\%.  Among simulations that lost a single planet, the most likely destruction pathways were collision of a planet with the Sun (55\%; 92\% of the time this was Mercury); impacts between two planets (30\%; Venus was involved 86\% of the time); and ejections (15\%; Uranus or Neptune were the ejected planet in 92\% of cases).

Could the Solar System's orbital architecture -- specifically, Mercury's and the ice giants' survival, and the absence of planets interior to Mercury -- constrain the properties of the Sun's birth cluster?  We do not think so, because the Solar System's orbital architecture during the stellar birth cluster phase was different than the present-day one. The typical timescale for the duration of the birth cluster is comparable to that of gas-dominated protoplanetary disks~\citep[e.g.][]{williams11,pfalzner14}. The terrestrial planets -- perhaps including Mercury -- may not have been completely formed during the cluster phase~\citep[e.g.][]{raymond14}.  In addition, the giant planets' orbital architecture was likely different, with secular resonances in different locations~\citep[see][]{raymond09c,brasser09}, because their dynamical instability had not yet taken place~\citep[e.g.][]{morby18}.

\begin{figure}
	\includegraphics[width=\columnwidth]{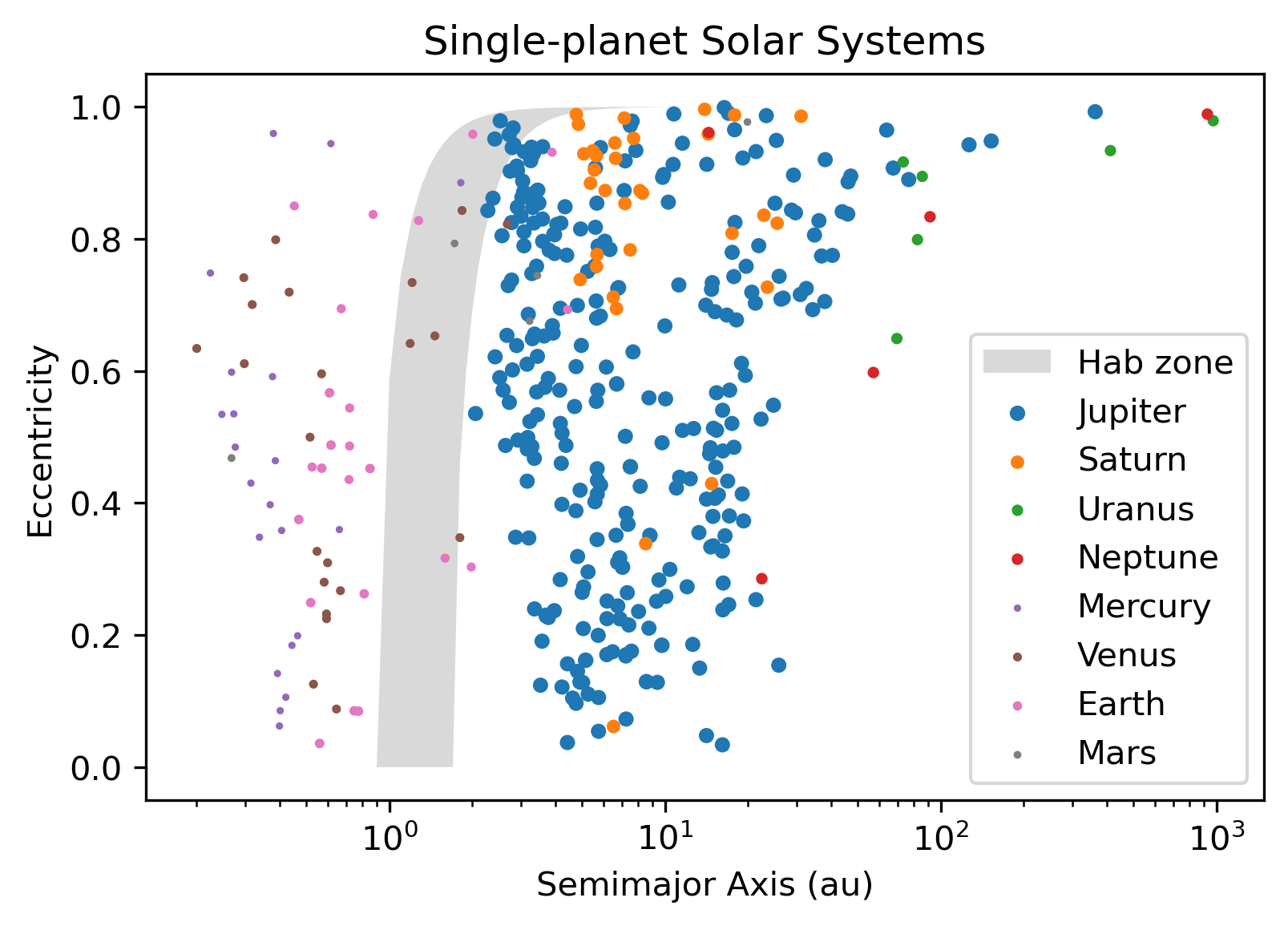}
    \caption{Final orbits of the 392 single-planet Solar Systems that were produced in our simulations.  The approximate boundaries of the habitable zone for the present-day Sun are shaded~\citep{kopparapu13}, included a scaling of the flux with orbital eccentricity~\citep{williams02,barnes09b}.   }
    \label{fig:singles}
\end{figure}

In 392 simulations, just one planet survived. Figure~\ref{fig:singles} shows the distribution of planetary orbits in single-planet Solar Systems.  Put together, a single planet has a total probability of $4.4\times 10^{-4}$ of surviving if a star passes within 100~au. Jupiter was the sole surviving planet in 278 of these cases, although there are simulations in which each planet is the lone survivor, including 23 solo-Earth systems (total probability of $1.06 \times 10^{-5}$). The surviving planet is rarely found in the habitable zone -- surviving terrestrial planets are almost universally found interior to the habitable zone and surviving giant planets exterior.  Nonetheless, there are a handful of habitable-zone planets, including a few systems with Jupiter on a very high eccentricity orbit -- this naturally provokes questions related to the habitability of Jupiter's Galilean moons~\citep[see][]{williams97} and whether they would survive system-wide dynamical instabilities~\citep[see][]{hong18}.  Given the very high eccentricities characteristic of single-planet systems, the climates of any habitable zone planets or moons would certainly undergo strong orbital variations~\citep[e.g.][]{williams02,dressing10,bolmont16}.

\subsection{Capture by the flyby star}

In some simulations, a planet was captured by the passing star.  We identified these cases, and re-ran the simulations with a much higher output frequency to capture the dynamics of the capture. 

\begin{figure}
 	\includegraphics[width=\columnwidth]{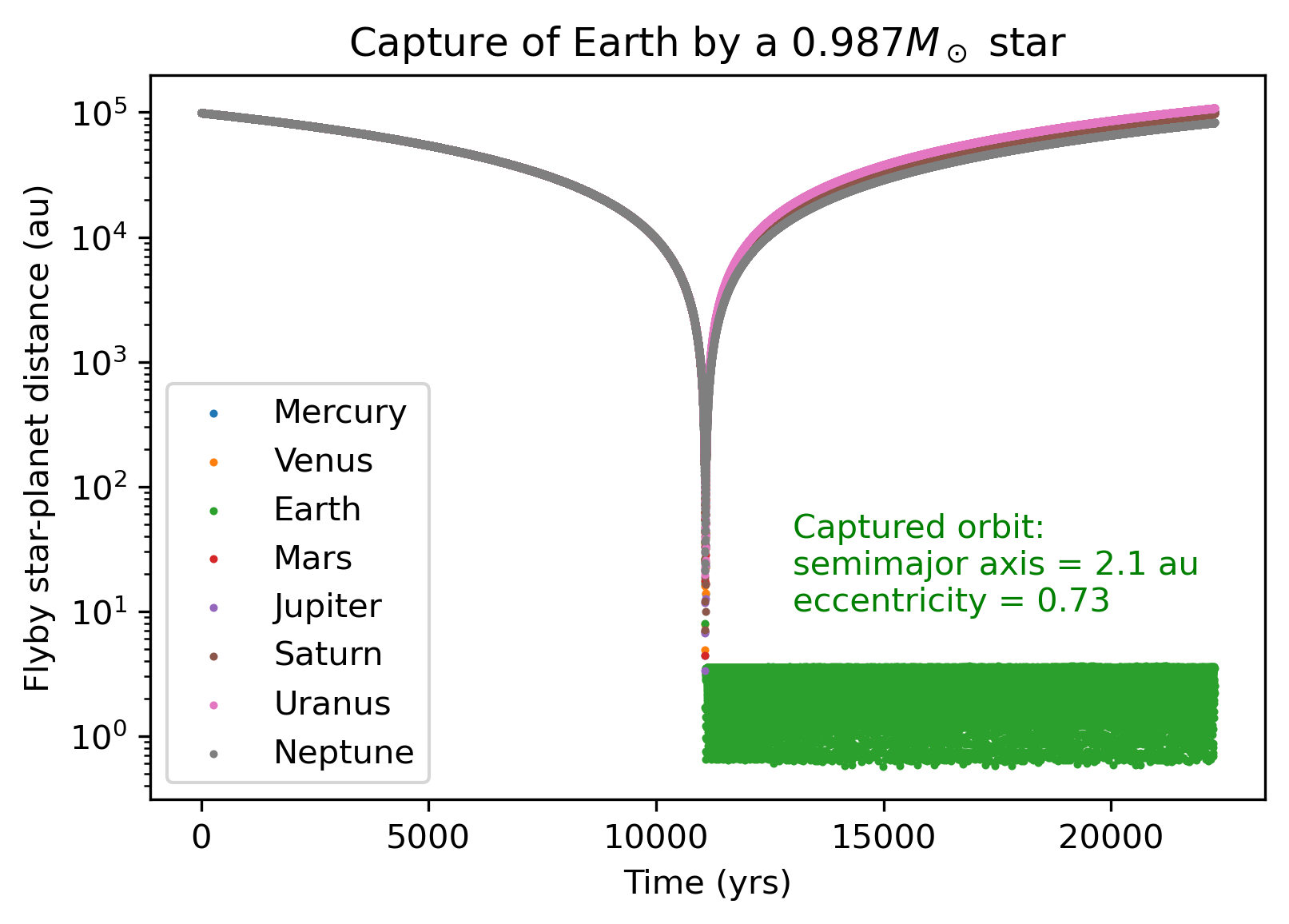}
    \caption{A simulation in which Earth was captured by a $\sim$~Solar-mass star onto a wider, eccentric orbit.  The vertical axis shows the distance between each planet and the encounter star, which passed close to the planets at $t \approx 11,000$~years, capturing only the Earth.}
    \label{fig:capture_e}
\end{figure}

Figure~\ref{fig:capture_e} shows a simulation in which Earth was captured during the flyby of a $0.987 M_\odot$ star with an impact parameter of 2.6~au at relatively low speed (21.4 km/s).  Jupiter was the lone surviving planet, ending up with an orbital semimajor axis of 11.6~au and an eccentricity of 0.94.  Saturn, Uranus and Neptune were ejected, while Mercury, Venus and Mars collided with the Sun.  Earth's new captured orbit was colder than its current one, with a semimajor axis of 2.1 au.  Its large orbital eccentricity of 0.73 would lead to strong climate variations over an orbit~\citep[e.g.][]{williams02,dressing10}.

Capture by another star is a low probability event. All together, 163 planets were captured by a flyby star in our simulations (including a handful of cases in which multiple planets were captured).  The total probability of any planet being captured if a star passes within 100~au is $1.6\times 10^{-4}$.  Saturn and Uranus had the highest capture probabilities, of $4 \times 10^{-5}$ and $4.6 \times 10^{-5}$, respectively.  Neptune and Mercury had the lowest capture probabilities, roughly an order of magnitude below those of Saturn and Uranus (see Table~\ref{tab:outcomes}).

\begin{figure}
	\includegraphics[width=\columnwidth]{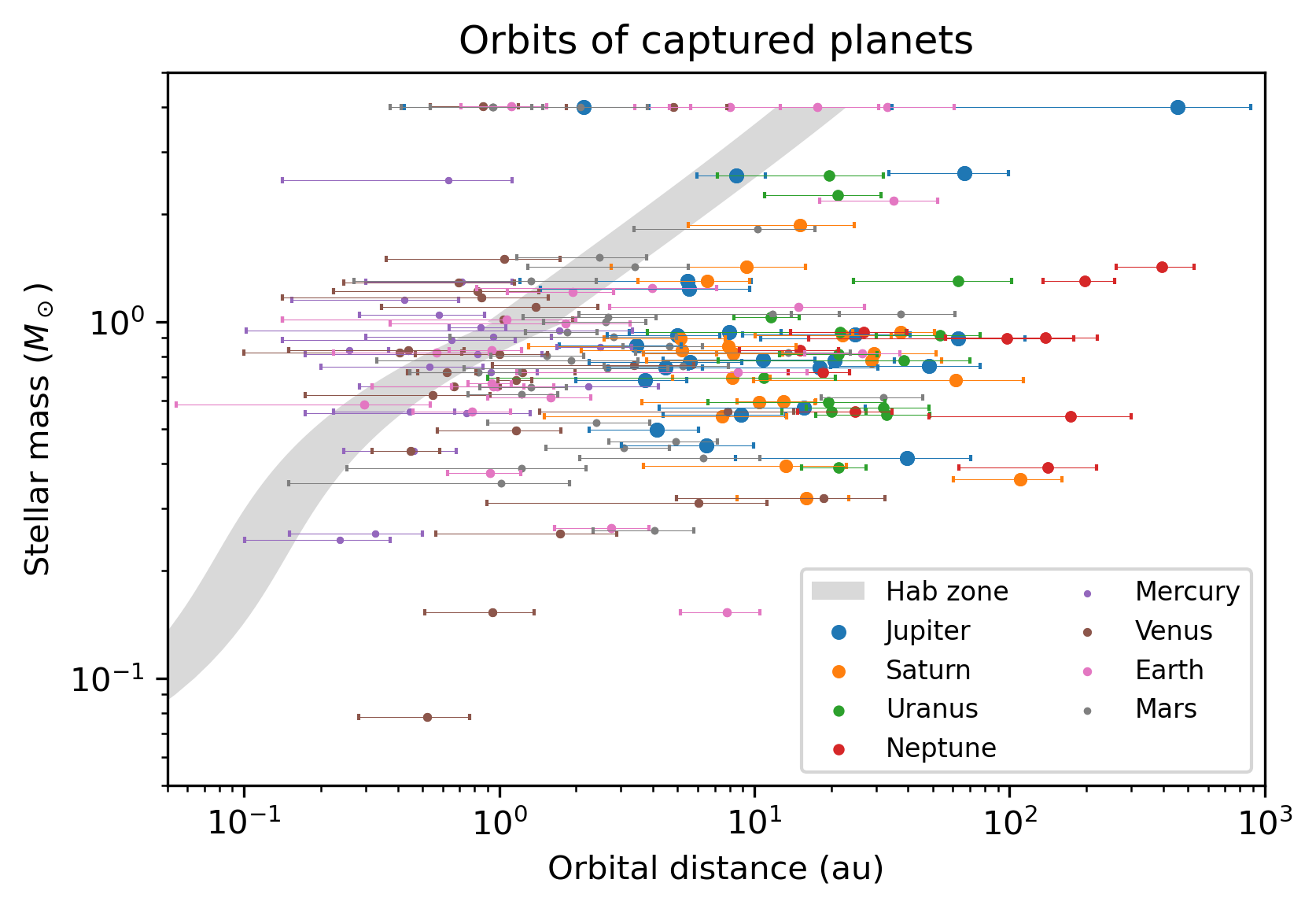}
    \caption{Orbits of planets captured by flyby stars.  For each planet, the horizontal error bar represents the radial excursion over its orbit due to orbital eccentricity. The habitable zone was estimated using the stellar mass-luminosity relation of \citet{scalo07} for low-mass stars and a simple $L\star \sim M_\star^4$ scaling for more massive stars, and assuming that the planetary albedo varied linearly from 5\% for the lowest-mass stars, to 30\% for Solar-mass stars, to 50\% for the highest-mass stars in our sample.}
    \label{fig:captures}
\end{figure}

Figure~\ref{fig:captures} shows the orbital radii of captured planets as a function of their new host star's mass. The orbits of captured planets are generally highly eccentric and follow a roughly thermal distribution with a median of $\sim 0.7$.  The orbital radii of captured planets around their new host stars are generally somewhat wider than their original values.  Still, many planets (generally the terrestrials) end up on orbits within their new host's habitable zone.

\begin{figure}
	\includegraphics[width=\columnwidth]{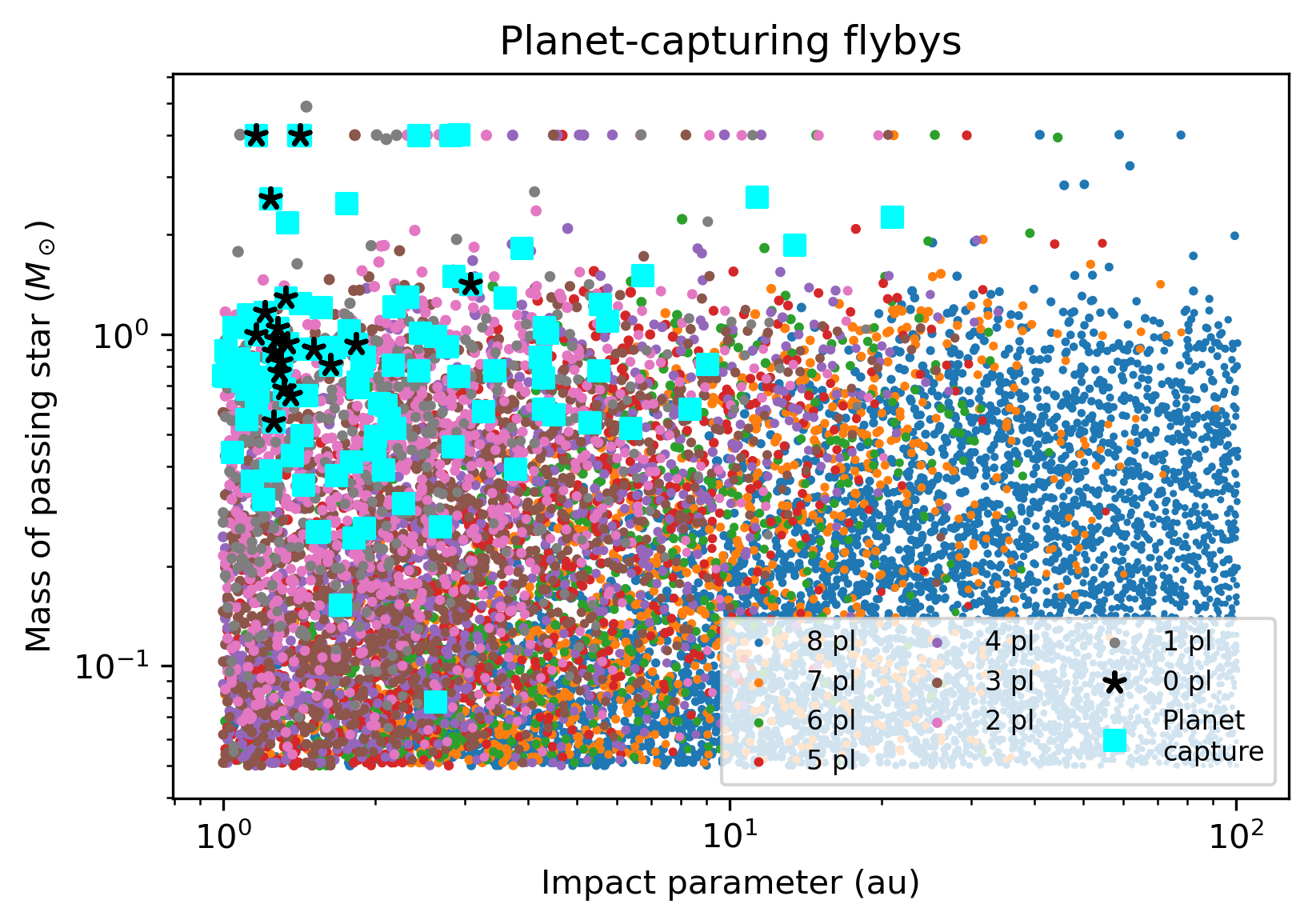}
    \caption{Characteristics of stellar flybys that led to the capture of a Solar System planet.  The size of each symbol is proportional to the log of the impulse gradient. }
    \label{fig:capture_mstar_b}
\end{figure}

Figure~\ref{fig:capture_mstar_b} shows that the encounters that captured planets were typically with high-mass stars moving at low relative velocity.  The median stellar mass among capture stars was $0.81 M_\odot$, almost five times higher than the median in all simulations of $0.16 M_\odot$.  The median stellar velocity (at infinity) was just 11 km/s, almost four times lower than the median in all simulations of 40.2 km/s. Put together, the median impulse gradient of planet-capturing encounters was $47$~km/s/au, 283 times higher than the median among simulations of 0.17~km/s/au.  It is interesting to note that systems that ended up with no surviving planets at all followed the same trends as planetary captures, with stellar masses (median of $0.94 M_\odot$), velocities (median of 12.6~km/s), and impulse gradients (median of 70.7 km/s/au).  This is not a surprise, as captures are a subset of outcomes of planetary stripping, in which the planet's liberated orbit happens to align with the perturbing star's. 

It is worth keeping in mind that our simulations assumed that the flyby star had no planets of its own.  We know from decades of exoplanets surveys that virtually all stars host at least one planet~\citep[e.g.][]{cassan12,fressin13,he19}, so this is statistically unlikely.  One can imagine Earth being captured by a new host star on an orbit that is friendly to life, only to undergo a sterilizing giant impact with another planet.  Alternately, the Sun could capture one or more planets from another star that end up wreaking havoc on the Solar System.  

\section{Simulations including the full Earth-Moon system}

In our main set of simulations, the Earth and Moon were treated as a single particle with their combined mass. To explore the importance of the Earth-Moon system, we re-ran a subset of our simulations including the Earth and Moon as separate particles.  We included them at their present-day orbital separation, maintaining their center of mass at the same position as the single particle in our other simulations. The orbits of all other planets were identical. 

We re-ran simulations in which, as a single Earth+Moon particle, Earth underwent the full range of outcomes, including ejection, collision with the Sun, impact with another planet, and regular, stable evolution.  We re-ran each simulation 12 times, varying the orbital phase of the Moon around the Earth in $30^\circ$ increments.  While this approach does not strictly capture the exact configuration of the Earth-Moon system relative to the planets' orbits, it allows us to get a statistical view of likely outcomes.  As a sanity check, we ran 12 simulations of the Solar System with no stellar flyby and the full range of Earth-Moon configurations.  All remained stable for 20 Myr.

\begin{figure}
	\includegraphics[width=\columnwidth]{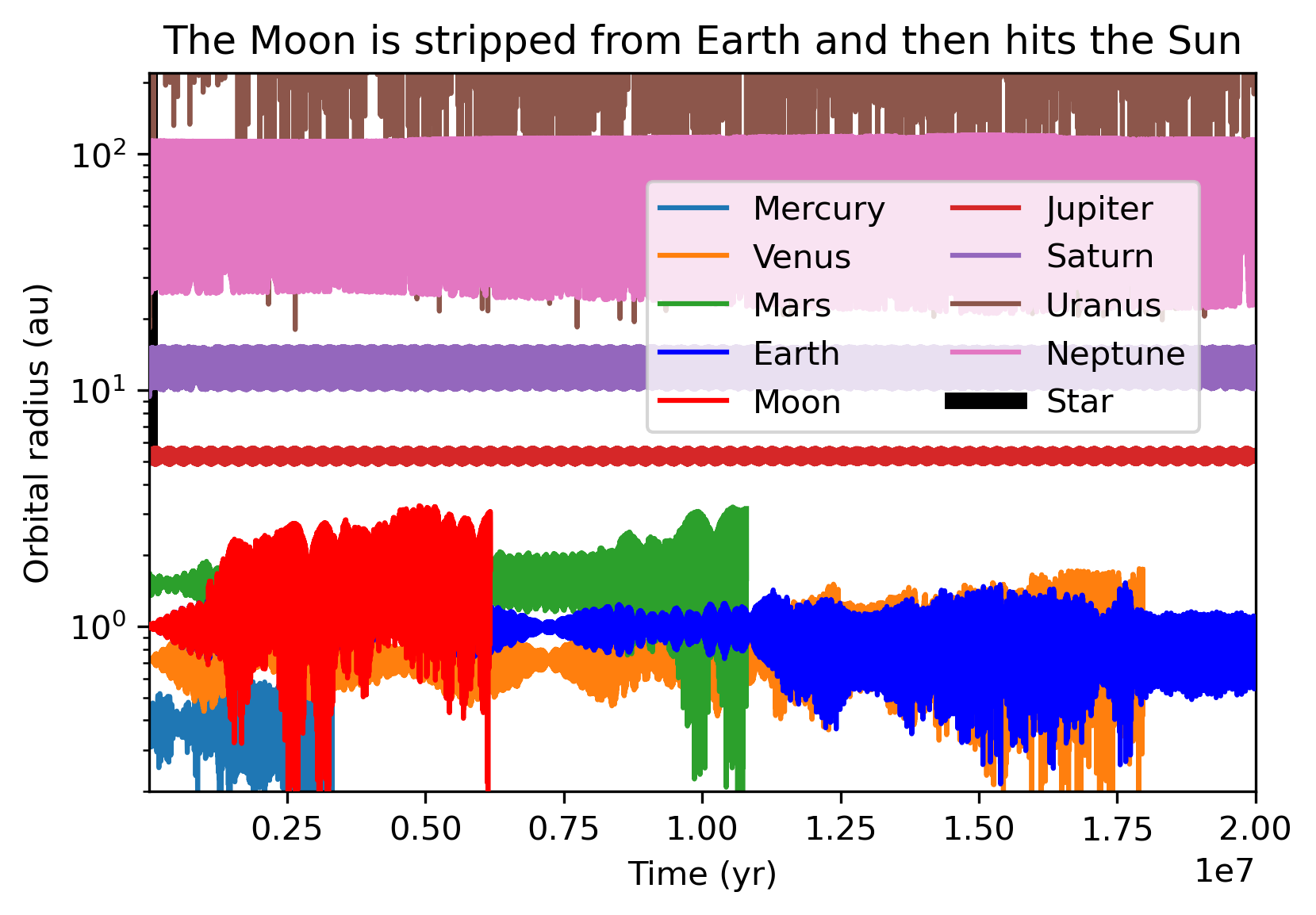}
    \caption{Evolution of a system in which the Earth and Moon were each included as separate particles. In this case, the Moon was stripped from the Earth after a close encounter with Venus, and later collided with the Sun.  Earth survived as a Moon-less world.}
    \label{fig:ae_moon}
\end{figure}

Figure~\ref{fig:ae_moon} shows a simulation in which Earth survived without a Moon.  First, a close stellar flyby (impact parameter = 6.2~au) triggered a dynamical instability that quickly spread through the whole Solar System. The terrestrial planets began a phase of strong interactions, and Mercury collided with the Sun. The Moon was stripped from the Earth after a close encounter with Venus, and collided with the Sun shortly thereafter.  Mars and then Venus also collided with the Sun, and Earth survived on a closer-in, eccentric orbit ($a = 0.82$~au, $e \approx 0.3$).  

It is not surprising that the Moon's orbit should be destabilized during a dynamical instability.  The Moon orbits Earth at roughly half the maximum stable orbital distance of $\sim 0.5 R_H$, where $R_H$ is the Hill radius, defined as $R_H = a (m/3 M_\star)^{1/3}$, where $a$ is the semimajor axis, and $m$ and $M_\star$ are the planet and star's masses, respectively~\citep{domingos06}.  There is a decent chance of orbital disruption if a planet's close encounter distance is comparable to a satellite's orbital radius~\citep[e.g.][]{hong18}.  This can lead to the satellite colliding with the planet, modification of the satellite's orbit, or stripping of the satellite from the planet entirely.  In some cases, it can even lead to capture of the satellite~\citep[as is thought to have been the case for Neptune's moon Triton;][]{agnor06}.  

In total we re-ran 71 cases including the Earth and Moon as separate particles, for a total of 852 simulations (with 12 different Moon phases per case).  These cases were divided between five broad categories of outcomes, based on the outcome for the single-particle Earth:
\begin{itemize}
    \item Stable systems, in which all 8 planets remained on well-behaved orbits with low $AMD$.
    \item `Calm-ish' systems, in which one or more planets were destroyed, but Earth survived (although sometimes on a perturbed orbit, or even one that might eventually be destabilized; the simulation from Fig.~\ref{fig:ae_moon} is from the `calm-ish' batch.).
    \item Simulations in which Earth underwent a collision with another planet.
    \item Simulations in which Earth was ejected.
    \item Simulations in which Earth collided with the Sun.
\end{itemize}

\begin{figure}
	\includegraphics[width=\columnwidth]{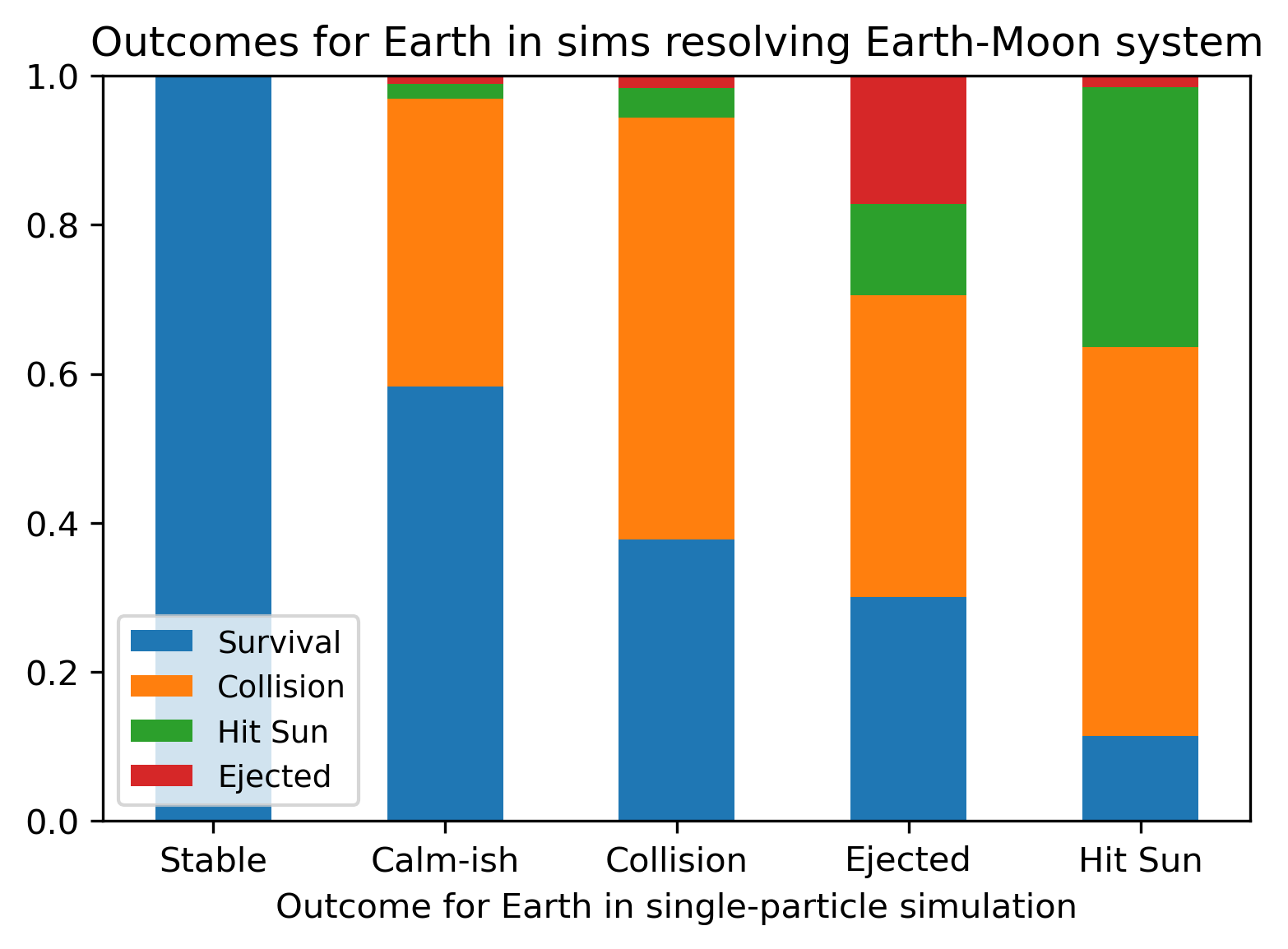}
    \caption{Outcomes for the Earth in simulations that were re-run including the full Earth-Moon system, labeled by the outcome in simulations with a single Earth+Moon particle.  Situations in which Earth first underwent a collision and later was ejected or hit the Sun were included as collisions (because that was the first process that would have destroyed the habitable conditions on Earth's surface). }
    \label{fig:outcomes_moon}
\end{figure}

Figure~\ref{fig:outcomes_moon} shows the outcomes for Earth in simulations that were re-run including the Earth and Moon as separate particles.  The outcomes were strongly dependent on the category of outcome.  All `Stable' simulations remained stable when the Earth and Moon were included as separate particles.

The presence of the Moon greatly increased the danger to Earth among the `Calm-ish' cases.  In more than 40\% of simulations, Earth either underwent a collision (with the Moon in 2/3 of impacts), was ejected, or hit the Sun.  In another 10\% of cases, the Moon was stripped from the Earth and usually collided with the Sun (or in one case, with Venus).
These cases -- in which a single-particle Earth escaped orbital disruption despite instability elsewhere in the system -- are those in which including the full Earth-Moon system makes the biggest difference.  This is because, compared with a single particle, the Earth-Moon system has a dramatically larger cross section for interaction with other planets that may roam through the inner Solar System. 

Among simulations that were more destructive for the Earth as a single particle, the outcomes were broadly similar.  Among cases in which Earth as a single particle suffered an impact with another planet (usually Venus or Mars), it still underwent a giant impact in 57\% of simulations.  However, the most common impactor was the Moon, which hit the Earth in 48\% of simulations (even though Earth subsequently was impacted by another planet in many cases).  

When considering cases in which Earth as a single particle was ejected from the Solar System, the outcome with the full Earth-Moon system depended on the mode of ejection.  When the ejection happened very early, triggered by the flyby star, the Earth and Moon were both ejected in all cases.  However, among systems in which Earth as a single particle had been ejected later, after a system-wide dynamical instability, less than 10\% of simulations led to the ejection of the Earth or Moon.  This was comparable to the ejection rate in the `Collision' and `Hit Sun' simulations. Rather, these simulations more often led to collisions with the Sun or with other planets.  This is because ejection of the Earth requires a series of gravitational scattering events leading to an increase in orbital energy.  Including the Earth and Moon as separate particles means that the exact series of events from the single-particle case cannot be reproduced. 

\begin{figure}
	\includegraphics[width=\columnwidth]{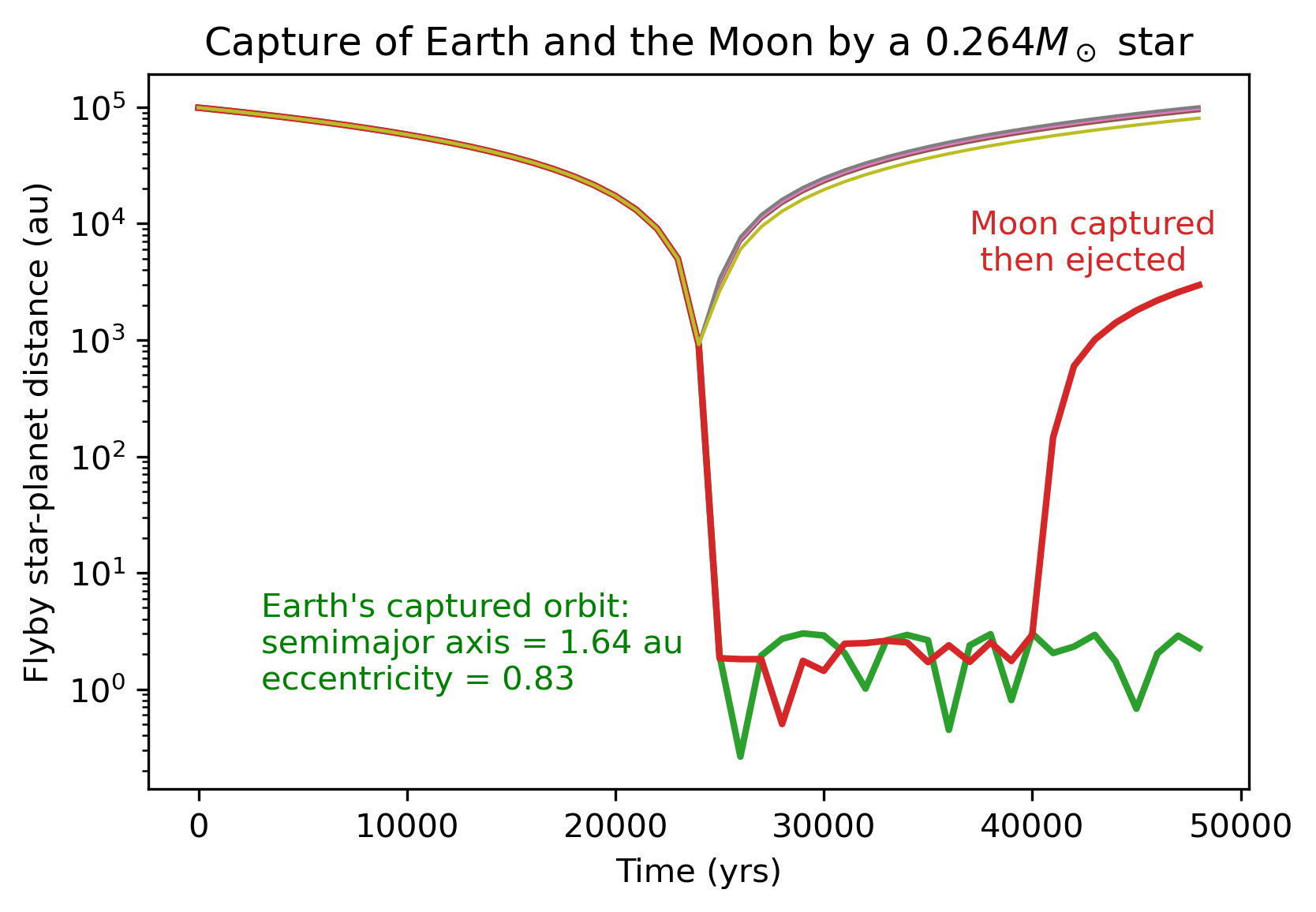}
    \caption{A simulation in which the Earth and Moon were both captured during the flyby of a low-mass star, but the Moon was later ejected.  }
    \label{fig:capture_moon}
\end{figure}

In principle, the Earth-Moon system can survive being ejected through gravitational scattering by a giant planet~\citep{debes07}.  However, among systems in which the Earth was ejected late, as a result of a system-wide dynamical instability, we only found a single case in which the Earth-Moon system survived (out of 11 ejections).  This is likely because of the large number of scattering events leading to their ejection.  Among system in which the Earth and Moon were both ejected early, due to the strong impulse from the flyby star, the Moon was retained in orbit around the Earth in almost every case.  

Among cases in which Earth as a single planet hit the Sun, Earth still hit the Sun in 59\% of simulations when the Earth and Moon were included as separate particles. However, almost half of the time, Earth underwent a giant impact (usually with the Moon) before hitting the Sun. We classified these outcomes as `collisions' in Fig.~\ref{fig:outcomes_moon}, since it was these collisions that first would have sterilized the Earth.  In 30 cases, Earth (and the Moon) remained on stable orbits, bound to the Sun and to each other, without strong enough perturbations to destabilize the system (at least during the 20 Myr integration).  

We also re-ran some simulations in which Earth was captured by the flyby star. The outcomes with the Earth and Moon as separate particles were system- and phase-dependent.  In some cases, all twelve simulations resulted in Earth's capture by the star; the Moon was also captured in more than half of simulations.\footnote{While there were many cases in which Earth was captured but the Moon was not, there were no simulations in which the Moon was captured without the Earth.} In other cases, Earth was only captured in a minority of simulations for a given flyby.  This is simply because capture depends on a precise alignment of velocity vectors, and these vary depending on the Moon's exact orbital phase.  However, even among systems in which both the Earth and Moon were captured by the flyby star, the Moon was always lost from the Earth and essentially became its own planet.  While we did not follow the long-term evolution of the orbits around the flyby star, the Moon was in all cases on an unstable orbit crossing that of the Earth.  We expect such systems to usually end in collision, or sometimes in ejection.

Figure~\ref{fig:capture_moon} shows a case in which the Earth and Moon were both captured by a $0.26 M_\odot$ star.  However, the Moon was no longer in orbit around Earth, and after a series of gravitational encounters, the Moon was ejected into interstellar space.  Earth survived on a much colder and more eccentric orbit than its present-day one.  



\section{Earth's future habitability}

\begin{figure}
	\includegraphics[width=\columnwidth]{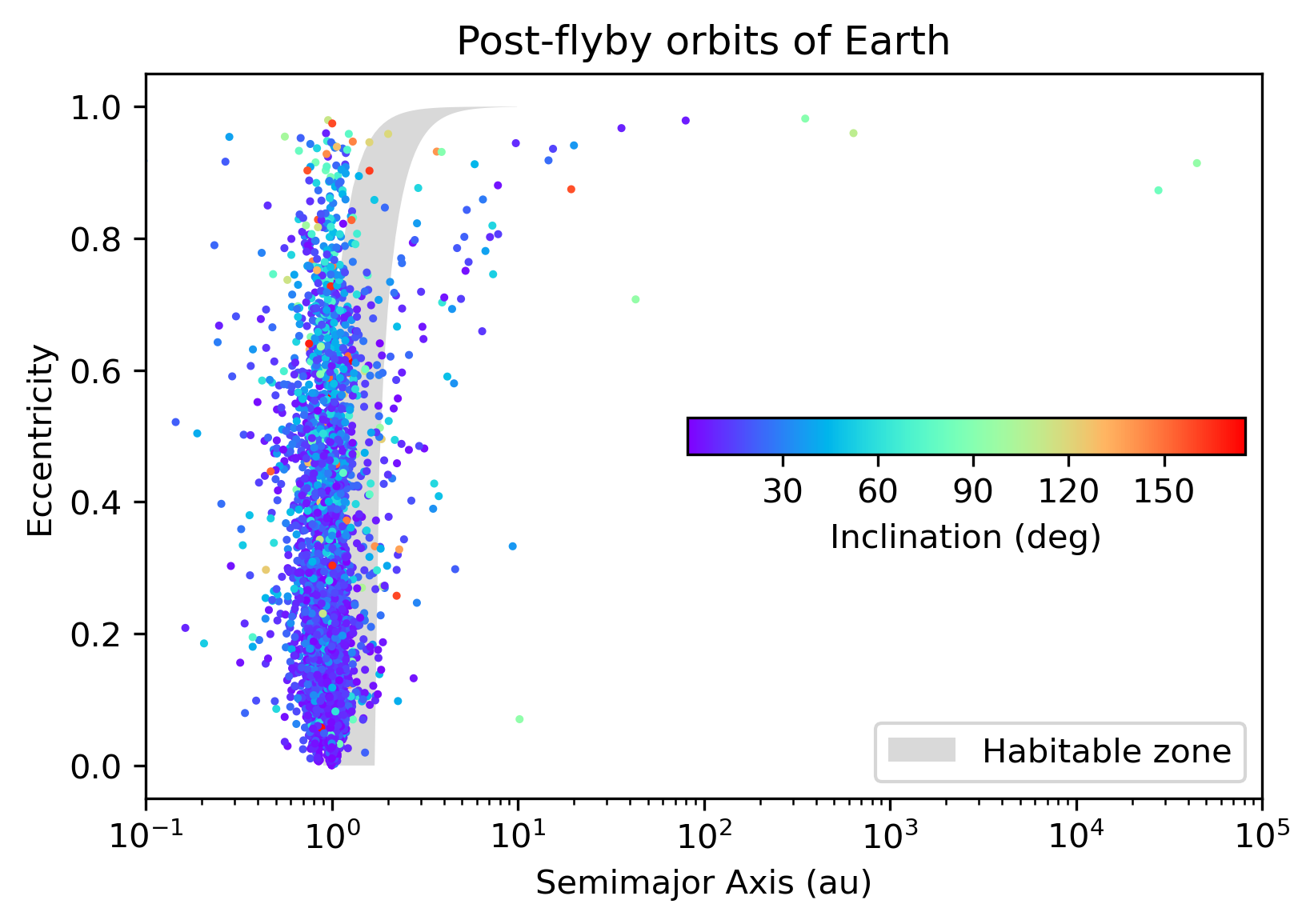}
    \caption{Post-flyby orbits of Earth around the Sun.  A handful of the high-semimajor axis orbits are unlikely to be stable on long timescales, but the two beyond $10^4$~au represent stable trapping in the Oort cloud -- the outermost case is the example from Fig.~\ref{fig:oort_e}.  The color of each planet corresponds to its inclination relative to Earth's original orbital plane. }
    \label{fig:ae_Earth}
\end{figure}

Figure~\ref{fig:ae_Earth} shows the distribution of post-flyby orbits of Earth, excluding cases in which Earth was captured around another star or ejected into interstellar space.  Earth's post-flyby semimajor axis tends to be close to its present-day one at 1~au, although scattering events can strand it closer or further from the Sun.  Of course, orbits far from 1~au tend to be very eccentric, although this is not a deal-breaker for habitability~\citep{williams02,barnes09b,dressing10,bolmont16}.

The surviving Earths in Fig.~\ref{fig:ae_Earth} have a range of orbital inclinations relative to Earth's original orbital plane.  While the probability of Earth surviving with a large inclination is small (see Fig.~\ref{fig:ei_dist}), this would certainly have an effect on Earth's climate, by changing its effective obliquity and, if other planets are present in the system, make it oscillate in time~\citep{spiegel09,raymond11,armstrong14}.

What are the odds that a stellar flyby could improve the prospects for the long-term future of life on Earth?  The best-case scenario is if Earth's post-flyby orbit is cooler than its present-day one. Unfortunately, a sub-100~au flyby only has a 0.28\% probability of making that happen.  We considered an orbit to be `cooler' if the incident Solar flux dropped to less than 90\% of the flux on a circular orbit at 1~au.  There was a substantially higher (0.79\%) chance of Earth's post-flyby orbit becoming hotter by at least 10\% in flux.  In the vast majority of cases ($>99\%$) there was no significant change in flux.

\begin{figure}
	\includegraphics[width=\columnwidth]{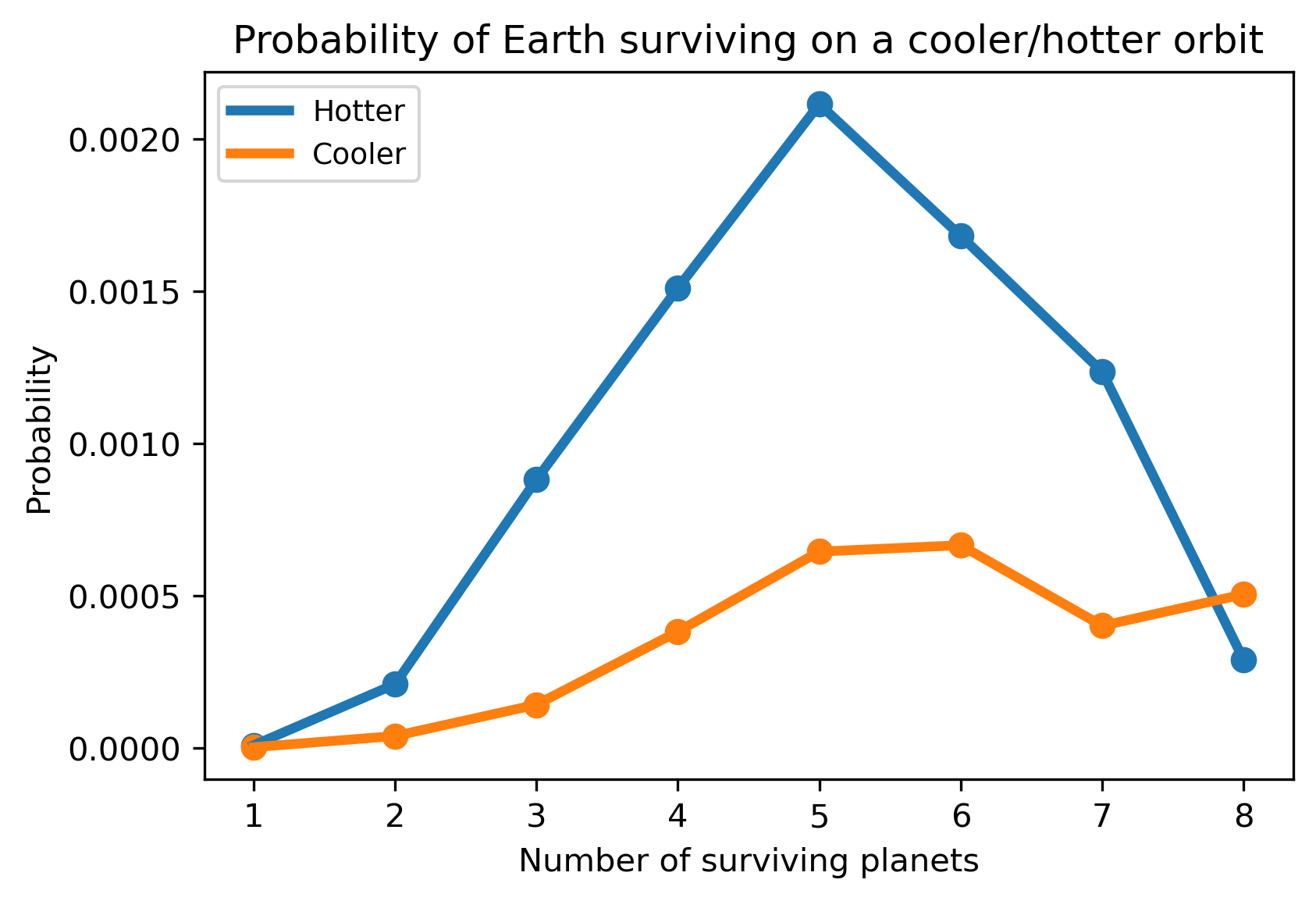}
    \caption{The probability that Earth survives on a hotter or cooler orbit as a function of the number of surviving planets, in the event of a sub-100~au stellar flyby.}
    \label{fig:cool_e}
\end{figure}

Figure~\ref{fig:cool_e} shows the probability of Earth surviving on a cooler (or hotter) orbit as a function of the number of surviving Solar System planets. There is a roughly comparable probability (of $\sim 0.0005$, or 1 part in 2000) of Earth ending up on a cooler orbit when the post-flyby Solar System contained 4, 5, 6, 7, or 8 planets. This takes into account a number of factors, including the probability of Earth surviving (including relative probabilities of different outcomes), and Earth's final orbit.  There was a much higher chance of Earth ending up on a cool orbit among systems with 4-5 surviving planets than among systems with 8 surviving planets, but the odds of that happening at all was much lower, given the correlation between outcomes and the impulse gradient of the flyby star (Fig.~\ref{fig:impgrad}).  

Earth's typical eccentricity progressively increased in systems with fewer surviving planets, and was markedly higher in cases when Earth's orbit was cooler than its present-day one.  In systems in which 7-8 planets survived, Earth typically was perturbed onto a slightly wider orbit, with a median semimajor axis of 1.08~au and a median eccentricity of 0.10.  However, in systems in which fewer planets survived, cooler Earths tended to have significantly higher eccentricities, with a median value of 0.18, 0.21, and 0.27 for 6, 5, and 4 planets surviving, respectively. The typical eccentricities among Earths that ended up on hotter orbits were even higher, with median values of 0.19, 0.28, and 0.34 for 6, 5, and 4 planets surviving, respectively. These represent strongly-perturbed systems in which Earth's orbital radius was shifted (either inward or outward) by gravitational scattering with the other planets.  Such strong scattering events invariably increase the planets' eccentricities.



A plausible long-term prospect for life on Earth may be for our planet to be ejected into interstellar space. \cite{laughlin00} explored the idea of the frozen Earth, and modeled its long-term thermal evolution (without any assumptions related to the atmosphere).  They found that it would take roughly 1 Myr for Earth's surface to freeze over completely, although they note that life could continue to thrive in hydrothermal vents and in the deep subsurface.

The outer edge of the habitable zone may effectively extend to infinity if a planet like Earth has a thick enough thermal blanket in the form of a many-bar, hydrogen-rich atmosphere~\citep{stevenson99,pierrehumbert11} or a several km-thick layer of ice~\citep{abbot11}.  In such cases, the planet's internal heat flux can provide enough heat to maintain stable liquid water.  If a free-floating Earth also kept its moon (as was the case in many of our simulations; see Section 4), then tides could provide an additional source of heat, which would be much higher than the present-day tidal heat flux if the Moon's orbit was modestly perturbed during ejection~\citep{debes07}. 

Free-floating planets are known to be abundant.  To date, hundreds have been detected via direct imaging~\citep[e.g.][]{miretroig22} and gravitational microlensing~\citep{mroz17,mroz20,sumi23}, with masses down to almost Earth's mass.  If they possess sufficient water and adequate thermal blankets, some of these (or their moons) may represent viable habitats for life.

\section{Long-term stability of the Solar System}

Studies evaluating the dynamical stability of the Solar System date back hundreds of years~\citep[for a review, see][]{laskar12}.  The stability of a pair of planets can be evaluated analytically~\citep{marchal82} or numerically~\citep{gladman93}. The Angular Momentum Deficit $AMD$ can itself be used as a proxy for dynamical stability of a pair of planets, given that it controls that maximum eccentricity that each planet can obtain and therefore the possibility of the two planets' orbits crossing~\citep{laskar17,petit18}. 

\begin{figure}
	\includegraphics[width=\columnwidth]{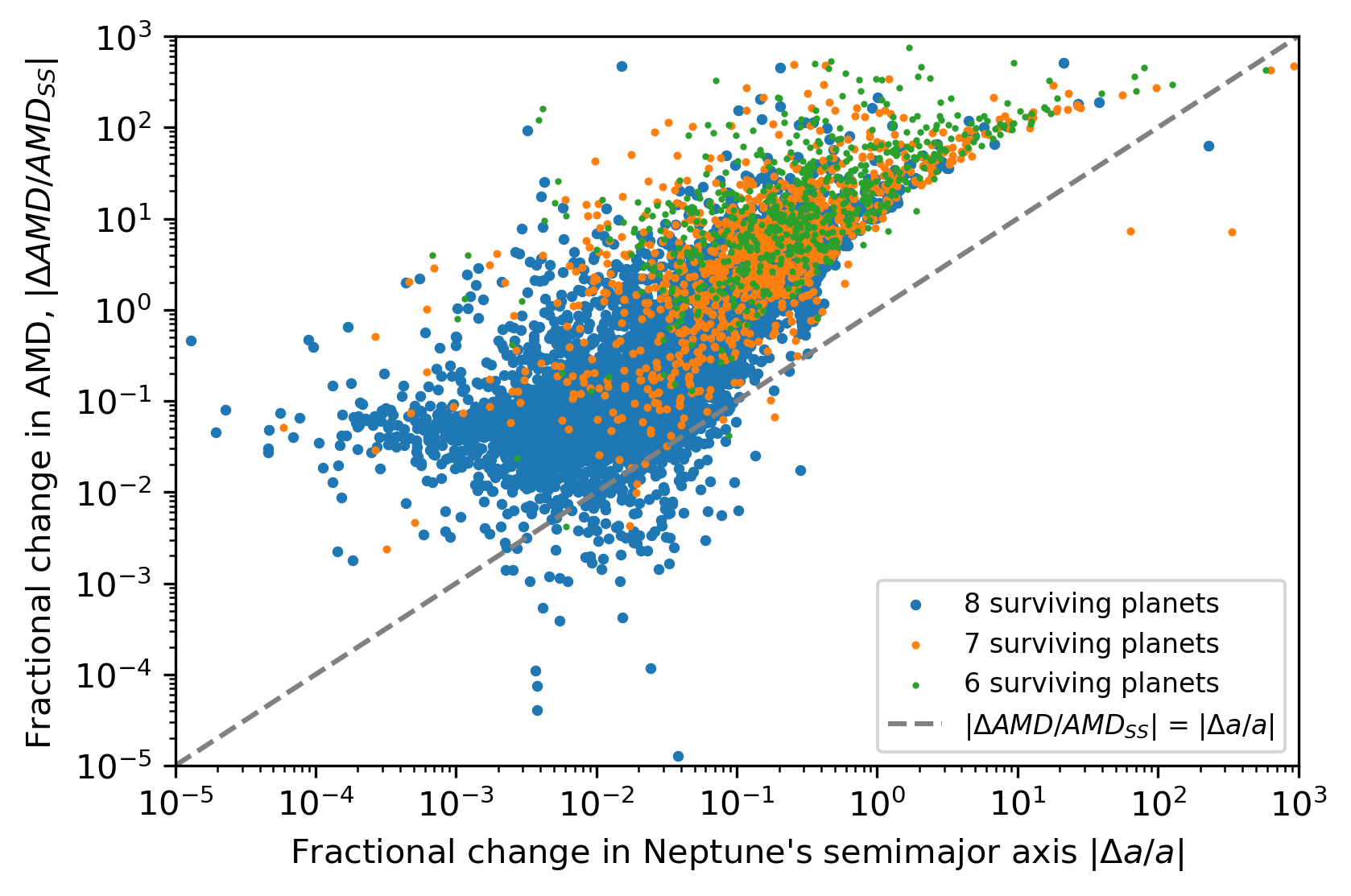}
    \caption{Fractional change in the angular momentum deficit of final 8-, 7- and 6-planet Solar Systems as a function of the fractional change in Neptune's orbit.}
    \label{fig:daa_nep}
\end{figure}

\cite{brown22} simulated the long-term stability of the Solar System under the influence of weak, distant stellar flybys.  They found that the probability of the system becoming unstable increased drastically when Neptune's semimajor axis was perturbed by more than 1 part in a thousand (roughly 0.03~au). Figure~\ref{fig:daa_nep} shows the fractional change in the angular momentum deficit $AMD$ of our systems with 6, 7 or 8 surviving planets as a function of the fractional change in Neptune's semimajor axis (calculated using Jacobi coordinates, to maintain maximum stability in semimajor axis).  We see the same general behavior as in the simulations of \cite{brown22}, with a strong trend toward higher $AMD$ in systems with a larger change in Neptune's semimajor axis.  This same trend is seen both within sets of simulations with 6, 7, or 8 surviving planets, and between the sets themselves.  This essentially boils down to the same trend that we showed in Fig~\ref{fig:impgrad}: stronger stellar perturbations (which have a higher impulse gradient, and which lead to larger changes in Neptune's semimajor axis) lead to more strongly disrupted systems.

\section{Discussion}

\subsection{Comparison with Previous work}


To our knowledge, the paper that is most similar in scope to our own is \cite{laughlin00}, who simulated the passage of binary stars by the Solar System and evaluated the probability of different outcomes. Our paper differs from \cite{laughlin00} in several ways: a) they used only binary flyby stars, while we used only single flyby stars (see discussion in Section 2.1); b) our flybys followed a logarithmic distribution in impact parameter $b$ (which we accounted for in calculating probabilities; see Section 2.3), while theirs followed a physically-motivated $b^2$ distribution; c) we included a prescription to account for the Galactic tidal field and therefore allow for the capture of planets into the Oort cloud; d) we included the full 8 planets in all of our simulations and ran them for 20 Myr, whereas, due to computational limitations, \cite{laughlin00} only included a subset of planets in most integrations, which only lasted 1 Myr; and e) we ran a subset of simulations that also included the Moon (see Section 5).  

Our results roughly match those of \cite{laughlin00} in terms of the different outcomes and general trends.  Of course, there are some outcomes that were out of reach for our simulations; for instance, given that our flyby stars were single, it was impossible for them to be captured by the Sun, which happened in the simulations of \cite{laughlin00}. Likewise, some of our outcomes were out of reach for \cite{laughlin00}, such as capture in the Oort cloud and interactions that involved the Moon.  

There are specific outcomes (such as those from Table~1) that we can compare, to gauge how closely our results match those of \cite{laughlin00}. They found a probability of 1 in 400,000 that Earth would be ejected from the Solar System in the next 3.5 Gyr, which amounts to $7.1 \times 10^{-7}$ per Gyr.  Our simulations yielded a significantly higher probability of $5.8 \times 10^{-6}$ per Gyr.  They found the odds that Earth is captured by a passing star to be roughly 1 in 2 million, or $1.4 \times 10^{-7}$ per Gyr, which is very close to our calculated value of $1.1 \times 10^{-7}$ per Gyr.  \cite{laughlin00} found a probability of $\sim 9 \times 10^{-6}$ per Gyr that Earth's post-flyby eccentricity would be larger than 0.5. We found a somewhat higher value of $2.4 \times 10^{-5}$ per Gyr.  Despite these modest differences, we consider our results to be broadly consistent.

\subsection{Limitations}

Our numerical simulations are admittedly imperfect on several fronts.  First, we did not include the gravitational influence of all Solar System bodies.  \cite{laskar11} showed that the largest asteroids provide a non-negligible contribution to chaos among the terrestrial planets on a $\sim 60$~Myr timescale.  Given the relatively short duration of our simulations and the Gyr or longer timescale for chaos to affect the stability of the terrestrial planets' orbits~\citep{laskar09,zeebe15,abbot21}, it seems unlikely that neglecting asteroids has a significant impact on the stability of the planets' orbits.  Yet we can imagine the possibility of asteroid or comet showers triggered by the destabilization of small body orbits during or after a flyby.

Second, as discussed above, we did not include binaries among the encounter stars.  Multiple-star systems are extremely common~\citep{duquennoy91}, and binary encounters can lead to certain interesting outcomes such as stellar capture~\citep[see][]{laughlin00}, which we could not capture in our simulations.  \cite{li15b} showed that the flyby of a binary can be approximated as simply the flyby of two single stars for encounter speeds faster than $\sim 20$~km/s. Many of the encounters that resulted in capture or complete planet stripping were at low encounter speed.  The cross section for interaction of a binary is more than twice that of a single star at low speeds~\citep{li15b}, which implies that our results may somewhat underestimate the probability of these outcomes.

Third, we did not include the full Galactic environment, but only a single close encounter with one passing star, and a simple recipe for torques from the Galactic tidal field.  Studies such as \cite{portegies21} developed a more consistent treatment for the Galactic environment, which does indeed affect the outskirts of planetary systems.  On a similar note, we only included a simplified approach for general relativity~\citep[taken from][]{saha92}, although other approaches exist~\citep[see][]{brown23}.

On a similar note, our simulations are limited to the conditions of the Sun's Galactic neighborhood.  The stellar number densities and velocity dispersions each vary by orders of magnitude among different Galactic locations such as globular or open (or embedded) star clusters, and the Galactic Bulge and core~\citep[see Table 1 of][]{brown22}.  The frequency of stellar flybys and their impact on planetary systems (via the impulse gradients) will vary accordingly.  The Solar System has experienced at least two different Galactic environments: its birth cluster and the local neighborhood.  It would be interesting to take the Sun's full Galactic history into account when assessing the stability of the Solar System.

\section{Conclusions}

In this paper we evaluated the dynamical evolution of the Solar System if a passes within 100~au of the Sun.  There is a $\sim$~1\% chance of this happening in the next Gyr~\citep{zink20,brown22}, which is the same time frame for the Earth to leave the habitable zone due to the increasing Solar luminosity.  Each of our 12000 simulations simulated the flyby of a single star within 100~au the Solar System.  We sample the impact parameter $b$ in a logarithmic fashion, allowing us to capture rare outcomes and still derive realistic probabilities (see Section 2.3).

Our main results are:
\begin{enumerate}
    \item The most likely outcome of a sub-100~au stellar flyby is for all 8 planets to survive on orbits similar to their present ones (see Fig.~\ref{fig:np_hist}).  There is a 92\% probability that the Solar System's angular momentum deficit (see Eq. 2) will also remain smaller than twice the current value with all pairs of neighboring planets on Hill stable orbits.
    \item The key parameter that controls the outcome is the impulse gradient, a measure of the strength of the perturbation from the stellar flyby (see Fig.~\ref{fig:impgrad} and Eq. 1).
    \item The post-flyby orbits of surviving planets can be very different than their present-day ones, in some cases with very high eccentricities and inclinations (Fig.~\ref{fig:ae_all}).  While extremely-excited orbits are a low-probability outcome  (Fig.~\ref{fig:ei_dist}), they would have consequences for planetary climate.
    \item If a star passes within 100~au of the Sun, the most likely planetary destruction pathways are: Mercury hitting the Sun (2.5\% probability), Mars colliding with the Sun (1.2\%), Venus colliding with another planet (likely Earth or Mercury; 1.2\%), or the ejection of Uranus (1.1\%) or Neptune (0.8\%).  Table~1 lists probabilities for all major outcomes for each planet.
    \item One of the most dramatic outcomes were simulations in which all 8 planets were stripped from the Solar System by a particularly strong stellar encounter (see bottom-right panel in Fig.~\ref{fig:examples}).  Luckily, the probability of this happening is just $8 \times 10^{-8}$ per Gyr. 
    \item In some simulations, a planet became trapped in the Sun's Oort cloud.  This is most likely to happen to the ice giants or Saturn (see Table~1), although Earth and Mars were each trapped in the Oort cloud in certain simulations. 
    \item In some simulations, one or more planets was captured into orbit around the flyby star (see Figs.~\ref{fig:capture_e} and~\ref{fig:captures}).  The eccentricities of captured planets tend to be very high.  Of course, most flyby stars would likely have their own planetary systems, which means that a captured planet may well undergo a phase of collisions -- or at least strong dynamical interactions -- with the flyby star's native planets (or, for that matter, with other captured planets; see Fig.~\ref{fig:capture_moon}). 
    \item Including the Moon as a separate particle (rather than having a single Earth-Moon particle with the sum of their masses) makes no apparent difference in the stability of the Solar System.  However, the presence of the Moon does open a window of new outcomes, the most common and destructive of which is a collision with the Earth (see Fig.~\ref{fig:outcomes_moon} and Section 4). 
    \item A stellar flyby is unlikely to allow Earth to remain habitable beyond a horizon of $\sim 1$~Gyr.  In the event of a sub-100~au flyby, there is only a 0.28\% probability that Earth will survive on an orbit that receives less than 90\% of the present-day stellar flux. The odds of Earth being ejected into interstellar space are even smaller, just $6 \times 10^{-6}$ in the next Gyr. If Earth is indeed ejected -- or captured on a wide orbit around a low-luminosity star -- the surface will freeze over completely on a timescale of roughly 1 million years~\citep{laughlin00}.
    \item Regarding the long-term survival of the Solar System, our simulations are consistent with the work of \cite{brown22}.  We see a clear correlation between the fractional change in $AMD$ and the fractional change in Neptune's semimajor axis (Fig.~\ref{fig:daa_nep}), although we argue that the impulse gradient is really the key factor controlling the outcome.
\end{enumerate}

Despite the diversity of potential evolutionary pathways, odds are high that our Solar System's current situation will not change.  Earth's progressive heating from the brightening Solar luminosity (not to mention human-driven carbon emissions) will continue unabated.  The Universe is statistically unlikely to help us out by providing a stellar flyby that will lead to a cooler Earth. Humanity's best solution is to help itself.

\section*{Acknowledgements}

The eventuality for the Earth to be ejected and its possible fates after such event initially came from a discussion between F.~S and the author luvan (\url{https://www.luvan.org/luvan-grav/about}), who is currently working on a science fiction verse novel. This research has received funding from the European Research Council (ERC) under the European Union’s Horizon 2020 research and innovation programme (grant agreement No 682903, P.I. H. Bouy), and from the French State in the framework of the "Investments for the future" Program, IdEx Bordeaux, reference ANR-10-IDEX-03-02.  S.~N.~R. and F.~S. acknowledge funding from the French Programme National de Planétologie (PNP) and in the framework of the Investments for the Future programme IdEx, Université de Bordeaux/RRI ORIGINS. N.~A.~K. acknowledges funding from NSF CAREER award 1846388 and NASA Exoplanets Research Program grant 80NSSC19K0445.

\section*{Data Availability}

All simulation data and analysis codes will be made available upon reasonable request.









\bsp	
\label{lastpage}
\end{document}